\documentclass[12]{article}

\def\square{\bullet} \def\NN{{\bf N}} \def\RR{{\bf R}}
\def\man{\mathcal{M}}
\def\beq{\begin{equation}}\def\eeq{\end{equation}}
\def\bea{\begin{eqnarray}}\def\eea{\end{eqnarray}}

\usepackage{epsfig}
\usepackage{color}
\def\half{{\textstyle{1\over2}}}\def\dd{{\rm d}}

\def\Blue#1{{\color{blue}#1}}

\definecolor{red}{rgb}{1,0,0}
\definecolor{blue}{rgb}{0,0,1}
\definecolor{green}{rgb}{0,1,0}
\definecolor{black}{rgb}{0,0,0}
\definecolor{yellow}{rgb}{1,1,0}
\definecolor{mdwblue}{rgb}{0.2,0.2,0.6}
\definecolor{gray}{rgb}{0.7,0.7,0.7}
\definecolor{darkgreen}{rgb}{0.2,0.7,0.2}
\author{Luca Bombelli$^1$ and Johan Noldus$^2$\\*\\
$^1$ Department of Physics and Astronomy, University of Mississippi\\
108 Lewis Hall, University, MS 38677, U.S.A.\\
$^2$ Vakgroep Wiskundige analyse, Galglaan 2, 9000 Gent, Belgium}

\begin{document}

\title{The moduli space of isometry classes\\
of globally hyperbolic spacetimes}
\maketitle
\begin{abstract}
\noindent This is the last article in a series of three initiated by the
second author.  We elaborate on the concepts and theorems constructed in
the previous articles. In particular, we prove that the GH and the GGH
uniformities previously introduced on the moduli space of isometry classes
of globally hyperbolic spacetimes are different, but the Cauchy sequences
which give rise to well-defined limit spaces coincide.  We then examine
properties of the strong metric introduced earlier on each spacetime, and
answer some questions concerning causality of limit spaces. Progress is
made towards a general definition of causality, and it is proven that
the GGH limit of a Cauchy sequence of $\mathcal{C}^{\pm}_{\alpha}$, path
metric Lorentz spaces is again a $\mathcal{C}^{\pm}_{\alpha}$, path metric
Lorentz space. Finally, we give a necessary and sufficient condition,
similar to the one of Gromov for the Riemannian case, for a class of
Lorentz spaces to be precompact.
\end{abstract}
\maketitle

\section{Introduction}

The geometry of individual spacetimes, modeled in classical general relativity
and similar theories by smooth manifolds with Lorentzian metrics, is a subject
that has been extensively studied for decades and is fairly well understood,
both locally and globally (see, for example, Ref \cite{Beem}); although
specific results may differ from those obtained in Riemannian geometry
\cite{Pete}, the field is also a well-developed one.  What is not nearly as
well developed is the study of the space of Lorentzian geometries, which
from the mathematical point of view includes questions about its topology,
metric structure, and the possibility of defining a measure on it, and from
the physics point of view is crucial for addressing questions such as when a
sequence of spacetimes converges to another spacetime, when two geometries
are close, or how to calculate an integral over all geometries.

To summarize what is known, we start by introducing a few concepts.  We will
denote by $\mathcal{L}(\man) = {\rm Lor}(\man)/\sim$ the space of all
Lorentzian geometries on a manifold $\man$, i.e., the space Lor$(\man)$ of
Lorentzian metrics on $\man$ modulo diffeomorphisms, and by $\mathcal{LS}$ the
much larger space of all Lorentz spaces (these definitions will be made more
precise below).  In this paper, we will consider cases in which the underlying
manifold $\man$ or space is compact, so we assume that to be the case from
now on. This implies, if the spacetime is to be regular and free of (almost)
closed timelike curves, the existence of spacelike boundaries or initial and
final hypersurfaces (we will therefore refer to these spacetimes as 
cobordisms).
Timelike boundaries may exist as well, but the spacetime can have closed
spatial sections instead.

Several topologies on $\mathcal{L}(\man)$ have been known for some time
\cite{Beem}, but distances on this space have been proposed only relatively
recently \cite{BS, Noldus0}.  The more interesting situation, however, is the
more general one without a fixed $\man$, and it turns out that the two
definitions of closeness that are known for that situation \cite{Bomb, Noldus1}
are also more interesting and more manageable even when used just on
$\mathcal{L}(\man)$.  Of the latter two proposals, the only one so far known
to give an actual distance function on $\mathcal{LS}$ is the one in Ref
\cite{Noldus1}, some of whose consequences were studied in Ref \cite{Noldus2};
this distance, and related concepts, are the tools we will use in this paper
to get a better understanding of the structure of the space $\mathcal{LS}$.

More specifically, in the next section we will recall the definitions of the
Gromov-Hausdorff distance $d_{\rm GH}$ between Lorentz spaces introduced in Ref
\cite{Noldus1}, as well a similar notion of closeness, and the Riemannian
distance $D$ (the ``strong metric") on each such Lorentz space $(\man,d)$
used in Ref \cite{Noldus2} (where this distance was denoted $D_\man$).
We will then use them to give a precise definition of Lorentz space and state
the questions we will address in the rest of the paper, where the power of the
strong metric will become clear.

\section{Basic definitions and the moduli space}

If $(\man,g)$ is a globally hyperbolic spacetime, the metric $g$ induces a
continuous Lorentzian distance on pairs $x,y\in\man$: $d_g(x,y)$ is the
supremum over all lengths of future oriented causal curves from $x$ to $y$,
if such curves exist, and zero otherwhise. Here we will generalize this
situation, and take the point of view that the primary objects are pairs
$(\man,d)$, where $\man$ is a set and $d$ a Lorentzian distance $d:
\man\times\man \to \RR^+\cup \{\infty\}$, satisfying (i) $d(x,x) = 0$,
(ii) $d(x,y) > 0$ implies $d(y,x) = 0$, and (iii) the ``reverse triangle
inequality" $d(x,z) \ge d(x,y) + d(y,z)$ for any $x,y,z\in \man$ such
that $d(x,y)\,d(y,z) > 0$.  Once such a pair $(\man,d)$ is given, a partial
order $\ll$ on $\man$, interpreted as a chronological relation between
events, can be quickly recovered by defining $x \ll y$ iff $d(x,y) > 0$; when
$\man$ is a manifold and $d$ continuous, this structure can be recovered from
a metric tensor as described above \cite{Beem}, but in general $d$ need not be
a Lorentzian path metric.\footnote{For a definition of Lorentzian path metric,
see Definition 7.}

The reason for emphasizing the use of pairs $(\man,d)$ to 
characterize spacetimes
here, rather than $(\man,g)$, is that they allow us to define \cite{Noldus1} a
Lorentzian version of the Gromov-Hausdorff distance \cite{Gromov} between
Riemannian manifolds. Specifically, given two spacetimes $(\man_1,d_1)$ and
$(\man_2,d_2)$, we define the Lorentzian Gromov-Hausdorff distance as
\beq
       d_{\rm GH}((\man_1,d_1),(\man_2,d_2))
       := \inf\{\epsilon \mid (\man_1,d_1)\ {\rm and}\ (\man_2,d_2)\ {\rm are}
       \ \epsilon{\rm-close} \}\;,
\eeq
where the two pairs are said to be $\epsilon$-close iff there exist two
mappings $\psi: \man_1 \to \man_2$ and $\zeta: \man_2 \to \man_1$ such that
for all $p_1,q_1 \in \man_1$ and $p_2,q_2 \in \man_2$,
\beq
       \big| d_2(\psi(p_1),\psi(q_1))-d_1(p_1,q_1) \big| \le \epsilon\;,\quad
       \big| d_1(\zeta(p_2),\zeta(q_2))-d_2(p_2,q_2) \big| \le \epsilon\;.
       \label{GH}
\eeq
The function $d_{\rm GH}$ is a distance on $\mathcal{L}(\man)$, so
$d_{\rm GH} ((\man_1,d_1), (\man_2,d_2)) = 0$ iff $(\man_1,d_1)$ and
$(\man_2,d_2)$ are diffeomorphism-equivalent.  However, well-defined limits for
Cauchy sequences of Lorentzian spaces have been obtained 
\cite{Noldus2} only with
a tighter definition of closeness, requiring that the mappings $\psi$ 
and $\zeta$
be approximate inverses of each other.  We say that $(\man_1,d_1)$ and
$(\man_2,d_2)$ are $(\epsilon,\delta)$-close iff there exist two mappings
$\psi$ and $\zeta$ as in (\ref{GH}), satisfying in addition
\bea
       \big| d_1(\zeta\circ\psi(p_1),q_1) + d_1(q_1,\zeta\circ\psi(p_1))
       - d_1(p_1,q_1) - d_1(q_1,p_1) \big| \le \delta\;, \nonumber\\
       \big| d_2(\psi\circ\zeta(p_2),q_2) + d_2(q_2,\psi\circ\zeta(p_2))
       - d_2(p_2,q_2) - d_2(q_2,p_2) \big| \le \delta\;,
\eea
for all $p_1,q_1 \in \man_1$ and $p_2,q_2 \in \man_2$. Such a definition of
closeness is captured by the mathematical notion of a uniformity; we call it
the Hausdorff (because it separates all points), quantitative (because of the
labels $(\epsilon,\delta)$), generalized Gromov-Hausdorff uniformity (GGH).

In the proof that Cauchy sequences in the GGH sense $\{(\man_i, 
d_i)\}_{i\in\NN}$
have well-defined limit spaces, an interesting tool emerged, a 
Riemannian (i.e.,
positive-definite) metric $D$, called {\em strong metric}, defined on each
$(\man, d)$ by
\beq
        D(p,q) = \max_{r \in \mathcal{M}}
        \big| d(p,r) + d(r,p) - d(q,r) - d(r,q) \big|\;. \label{D}
\eeq

The definitions and results summarized above, and in particular theorem 6 of
Ref \cite{Noldus2}, strongly suggest the following definition of a Lorentz
space.

\newtheorem{deffie}{Definition}
\begin{deffie}

A Lorentz space is a pair $(\man,d)$, where $\man$ is a set and $d$ is a
Lorentz distance on $\man$, such that $(\man, D)$ is a compact metric space.

\end{deffie}

\noindent We denote by $\aleph_{\rm c}$ the space of all such Lorentz spaces.
On $\aleph_{\rm c}$, we can introduce an equivalence relation $\sim$ 
by defining
$(\man_1, d_1) \sim (\man_2, d_2)$ iff there exists a bijection 
$\psi$ such that
$d_{2}(\psi(x), \psi(y)) = d_{1}(x,y)$ for all $x,y \in \man_1$. Such 
a bijection
is automatically a  homeomorphism, and therefore $\sim$ defines an equivalence
relation.

\begin{deffie}

The moduli space of all isometry classes of Lorentz spaces is the space
$\mathcal{LS} = \aleph _{c}/\sim$, equipped with the Hausdorff, quantitative,
generalized Gromov-Hausdorff uniformity.

\end{deffie}

\noindent It was shown in Ref \cite{Noldus2} that $\mathcal{LS}$ is a complete,
contractible space in which the finite  spaces form a dense subset. It is
easily seen that it is not locally compact.\\*
\\*
\textbf{Note}:
The results in section 3 of Ref \cite{Noldus1}, in particular theorem 6, imply
that the obvious extension of $d_{\rm GH}$ from $\mathcal{L}(\man)$ 
to the moduli
space of isometry classes of Lorentz spaces is also a metric.  However, in the
above definition, we prefer to equip this space with the GGH uniformity, since
$\mathcal{LS}$ is then complete.\footnote{The authors are unaware of any proof
of, or counterexample to the stement that the moduli space of isometry classes
equipped with $d_{\rm GH}$ is complete.} \hfill$\square$
\\*
\\*
Let us comment now a bit on why these results are so easy and completely
analogous to the Riemannian case.  In our personal opinion, things became
a lot easier than in previous attempts at defining distances between
Lorentzian spaces \cite{BS,Noldus0,Bomb} because we have quit considering the
causal relation, separately from the volume element, as being the fundamental
object.  This opened up the possibility of introducing the strong metric $D$,
which emerged despite its strong nonlocality as a natural object, and 
allowed us to
ask questions concerning closeness and convergence in a more direct 
and quantitative
manner.  The theorems in Ref \cite{Noldus2} clearly bring to light 
the technical
potential of the metric $D$.  On the other hand, the examples in that 
same paper
show that defining a suitable causal structure out of the 
chronological one might
prove to be a nontrivial task in the context of general Lorentz spaces, due to
the existence of degenerate regions in the limit spaces.  Moreover, 
causal curves
and causal relations (as opposed to chronological ones) in Lorentz spaces have
different properties from the ones which we are used to with globally 
hyperbolic
cobordisms.  For example, geodesics between two timelike related points are
not necessarily timelike curves.

The goal of this paper is to further study the properties of the moduli
space of Lorentzian geometries and the structures described above.  In
particular, we (1) Try to find out if the Lorentzian Gromov-Hausdorff metric
and the generalized Gromov-Hausdoff uniformity are equivalent in the sense that
they have the same Cauchy sequences; (2) Study the question whether the strong
metric determines the Lorentzian distance uniquely up to time reversal (this
would be particulary interesting, since if it were true, then Lorentzian
cobordisms would be a subclass of Riemannian, non-path compact metric spaces
modulo ${\bf Z}_2$) (3) Study the definition of a suitable causal relation and
causal curves on limit spaces of compact globally hyperbolic cobordisms (for
example, if we knew how to define a causal relation between two points in the
``degenerate area" of a limit space, then one could raise the question of
the physical meaning of such ``causal relationships"); (4) Deal with the
moduli space and some matters of precompactness.

\section{$d_{\rm GH}$ versus the generalized Gromov-Hausdorff uniformity}

In this section, we examine the relationship between the  Gromov-Hausdorff (GH)
distance and the generalized Gromov-Hausdorff (GGH) uniformity for Lorentz
spaces $(\mathcal{M},d)$.  Along this study, some questions raised in Ref
\cite{Noldus1} will be solved. Since the difference between GH-closeness and
GGH-closeness lies in the condition that the mappings used in the definition be
approximate inverses of each other, we find it useful to introduce the concepts
of approximate isometry and approximately surjective mapping.

\begin{deffie}

Given a Lorentz space $(\man,d)$ and an $\epsilon > 0$, a mapping
$f:\mathcal{M} \rightarrow \mathcal{M}$ is:

\begin{itemize}

\item an $\epsilon$-isometry iff $f$ changes $d$-distances between any pair of
points by no more than $\epsilon$; i.e., for all $x,y \in \man$
$$
        \big| d(f(x),f(y)) - d(x,y)  \big| < \epsilon \;;
$$
\item an $\epsilon$-surjection iff any point is within a $D$-distance
$\epsilon$ of the image of some point; i.e., for all
$p
\in
\mathcal{M}$ there exists a $q \in \man$ such that
$$
        D(p, f(q)) < \epsilon \;,
$$
where $D$ is the strong metric on $\mathcal M$ constructed using $(\man,d)$.
\end{itemize}

\end{deffie}
We start with the following theorem.
\newtheorem{theo}{Theorem}

\begin{theo}

Let $(\mathcal{M},g)$ be a compact, globally hyperbolic cobordism. Then,
for every $\eta > 0$ there exists an $\epsilon > 0$ such that, for any
$\epsilon$-isometry $f$, there is an isometry $h$ of $\man$ such that:
$$
        D(f(x), h(x)) < \eta \quad \forall x \in \mathcal{M}\;.
$$
\end{theo}

\noindent\textsl{Proof:}
\\*
\noindent Suppose that the statement is false.  Then, given $\eta > 0$,
for each $n \in \NN$ there exists a $\frac{1}{n}$-isometry $f_{n}$, such
that for any isometry $f$ we can find a point $x(n,f)$ in $\man$ such that
$$
        D(f_{n}(x(n,f)), f(x(n,f))) \geq \eta\;.
$$

The proof of theorem 6 in Ref \cite{Noldus1} reveals that the sequence
$\{f_n\}_{k\in\NN}$ has a subsequence $\{f_{n_{k}}\}_{k \in \NN}$ such that
$f_{n_{k}} \stackrel{k \rightarrow \infty}{\rightarrow} f$ pointwise.
We show now that this convergence is uniform in the strong metric,
which provides the necessary contradiction.  We restrict ourselves to
proving that for any interior point $p$ and $\epsilon >0$, there exists a
$\delta >0$ such that $q \in B_{D}(p, \delta)$ implies that $D(f(q),f_{n}(q))
<\epsilon$, for $n$ big enough; The rest of the statement is easy (but tedious)
and is left as an exercise to the courageous reader.  Choose two points
$s,r \in B_{D}(p, \frac{\epsilon}{2})$ such that $s \ll p \ll r$ with, say,
$d(s,p)=d(p,r)$ as large as possible.  Let $\delta = \frac{1}{3}d(p,r)$; then
for $n > \frac{1}{\delta}$ such that $f_{n}(r) \in B_{D}(f(r), \delta)$ and
$f_{n}(s) \in B_{D}(f(s), \delta)$ we have that
$$
     f_{n}(B_{D}(p, \delta)) \subset A(f(r),f(s)) \subset B_{D}(f(p),
     {\textstyle\frac{\epsilon}{2}})\;,
$$
where $A(\cdot,\cdot)$ denotes the Alexandrov set between two points in
$\man$. Hence,
$$
     D(f_{n}(q),f(q)) \leq \textstyle{\frac{\epsilon}{2}}+\delta < \epsilon\;,
$$
for all $q \in B_{D}(p, \delta)$. \hfill$\square$
\\*
\\*
This result reveals that for any $\eta > 0$, there exists an $\epsilon >0$
such that any $\epsilon$-isometry is an $\eta$-surjection.  This, however, is
a fairly weak result and we would like to know if $\eta$ could be bounded by
some universal function of $\epsilon$, the timelike diameter and dimension of
$\man$, which goes to zero when either $\epsilon$ or the timelike diameter
go to zero. An example will show that such a function cannot exist, and lead us
to other examples that address directly our goal for this section.

\newtheorem{example}{Example}

\begin{example}
\end{example}
Consider the region $\mathcal{R}_\delta = \{(x,t) \mid x \in [0,\pi]
,\ t \le T(x)\}$ of the 2-dimensional flat cylinder $\mathcal{CYL} = ({\rm S}^1
\times [0,1], -\dd t^{2} + \dd\theta^2)$, where $T$ is some differentiable
function satisfying $T(x) \geq  1 - \delta$ for all $x \in [0,\pi]$, $T(0)
= T(\pi) = 1$, $T({\pi\over2}) = 1-\delta$, and $|T'(x)| < 1$ (see Fig 1).
\begin{figure}[h]
	\begin{center}
		\scalebox{0.53}{\includegraphics{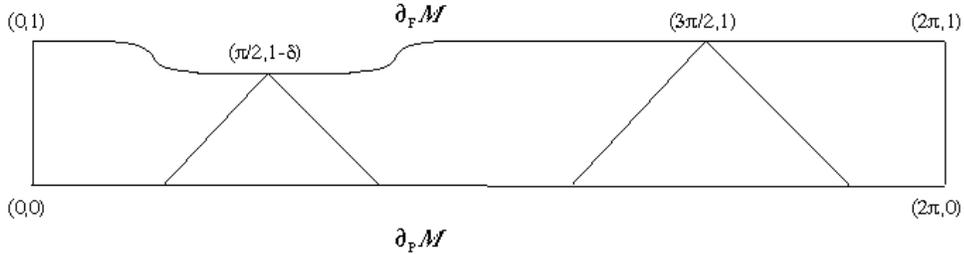}}
 	\end{center}
	\caption{Illustration of example 1}
	\label{fig:grliph5-1}
\end{figure}
We can now show an approximate isometry $\psi$ which is far from any
isometry.  The mapping $\psi$ is constructed as the composition of a rotation
by $\pi$ times a retraction $R_{\mathcal{R}_{\delta}}$ which maps a point
$(x,t)$ to the unique closest point $(x,\tilde{t}) \in \mathcal{R}_{\delta}$.
It is not difficult to verify that $\psi$ is a $\sqrt{2 \delta}$-isometry.
However, the point $p = (\frac{3\pi}{2},1)$ gets mapped to a point which is a
strong distance $1 = tdiam(\mathcal{R}_{\delta})$ away.\footnote{For a Lorentz
space $(\man,d)$ the timelike diameter is $tdiam(\man):= \max_{p,q\in\man}
d(p,q)$.}  This shows that for $\delta$ arbitrarily small, one can construct
spaces which allow $\delta$-isometries to be a distance $1$ apart from any
isometry (our analysis is simplified by the fact that the only isometry is the
identity!).
\hfill$\square$
\\*
\begin{example}
\end{example}
In this example, we show that a near isometry can be arbitrarily far from
being a surjection.  The following picture shows a sequence of $N$ ``bumps"
with a fixed width $L>1$.  Let $0 < \epsilon < \frac{1}{2}$ and consider the
function $g_{\epsilon} : [-\epsilon, \epsilon] \rightarrow
\RR^{+} : x \rightarrow \epsilon + x^{2}$.  Define a sequence of functions
$\Omega^{i}_{\epsilon} : \left[ (i-1)L , iL \right] \rightarrow \RR^{+}$, $i=1
\ldots N$, which satisfy the following properties:
\begin{itemize}
\item $ 0 \leq \Omega^{i+1}_{\epsilon}(x+L) -
\Omega^{i}_{\epsilon}(x) \leq \frac{L}{\sqrt{2}N}$ for $1 \le i \le N-1$ and
$x \in \left[ (i-1)L , iL \right]$.
\item $\Omega^{i}_{\epsilon}$ is symmetric around $x = (i-\frac{1}{2})L$.
\item $\max_{x \in \left[ (i-1)L , iL \right] }
\Omega^{i}_{\epsilon}(x) = i\,\frac{L}{\sqrt2N}$.
\item $\Omega^{i}_{\epsilon} (x) = g_{\epsilon}(x - (i-1)L)$ for all $x
\in \left[ (i-1)L, (i-1)L + \epsilon \right]$.
\item $ \left| \dd\Omega^{i}_{\epsilon}(x)/\dd x \right| < 1$ for
all $x \in \left[ (i-1)L , iL \right]$.
\end{itemize}
Let $\Omega_{\epsilon}$ be the concatenation of all the 
$\Omega^{i}_{\epsilon}$.
By identifying $0$ and $NL$, we obtain that $\Omega$ is a smooth function on
the circle of radius $NL/2\pi$.  Define $\mathcal{A}_\epsilon$ as
$$
     \mathcal{A}_\epsilon = \left\{ (x,t) \mid x \in [0,NL] \textrm{ and }
     t \in [0,\Omega_\epsilon(x)] \right\}.
$$
Then $(\mathcal{A}_\epsilon, -\dd t^2 + \dd x^2)$ is a globally hyperbolic
cobordism cut out of the cylinder universe $\mathcal{CYL} = {\rm S}^1 \times
\RR$ with radius $\frac{NL}{2 \pi}$.  Define $\psi : \mathcal{A}_\epsilon
\rightarrow \mathcal{A}_\epsilon$ as the composition of a rotation to the
left over an angle of $\frac{2 \pi}{N}$ with a retraction
$R_{\mathcal{A}_\epsilon}: \mathcal{CYL} \rightarrow \mathcal{CYL}$
which maps every point $(x,t)$ to the closest point $(x,\tilde{t}) \in
\mathcal{A}_\epsilon$.  Considering the point $((N-\half)L,{L\over\sqrt2})$ and
suitable other points, we clearly see that $\psi$ is
a $\frac{L}{\sqrt{N}}$-isometry which is not a $\frac{(N-1)L}{\sqrt{2}
N}$-surjection.  The figure will be called the carousel for obvious reasons.
\hfill$\square$
\begin{figure}[]
	\scalebox{0.45}{\includegraphics{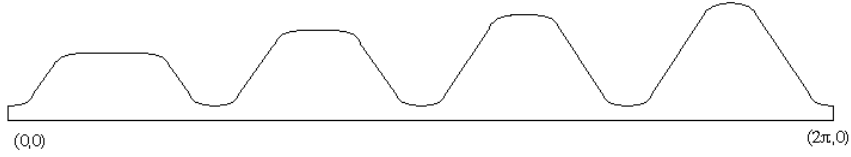}}
    \caption{The carousel}
	\label{fig:grliph5-2}
\end{figure}
\\*
\\*
Let $f,g : \mathcal{LS} \times \mathcal{LS} \times \RR^{+} \rightarrow
\RR^{+}$ be functions depending only upon the timelike diameters, $f(x,y, 0) =
g(x,y,0) = 0$ and $f,g$ are continuous in the third element in $(x,y,0)$ for
all $x,y \in \mathcal{LS}$.  Do functions satisfying the above conditions
exist such that if $x$ and $y$ are $\epsilon$-close then they are $(f(x,y,
\epsilon) , g(x,y, \epsilon))$-close?  It is not difficult to prove that if
there exist two mappings $\psi, \zeta$ which make $x$ and $y$ $\epsilon$-close
such and $\psi$ or $\zeta$ is \emph{surjective}, then $x$ and $y$ are
$(\epsilon, 2\epsilon)$-close.  Hence if we want to find a counterexample then
we have to look for mappings which are ``far" from being surjections.  The
carousel hints at the following counterexample.
\begin{example}
\end{example}
Suppose $L=4m$, $m \in \NN \setminus \left\{ 0,1 \right\}$ and let
$\mathcal{P}^{L}_{1}, \mathcal{P}^{L}_{2}$ be causal sets given by the Hasse
diagrams below, where it is understood that the points in each of the thicker
pairs are to be identified.  On a locally finite causal set $\mathcal{P}$, the
maximum number of links between two timelike related points $p \ll q$ gives a
Lorentz distance, and if $\mathcal{P}$ is finite it has a natural
interpretation as a Lorentz space.  Obviously $\mathcal{P}^{L}_{1},
\mathcal{P}^{L}_{2}$ are $1$-close, as one can see using a map
$\psi : \mathcal{P}^L_1 \to \mathcal{P}^L_2$ which takes each column of
$\mathcal{P}^L_1$ to the corresponding one in $\mathcal{P}^L_2$, and a map
$\zeta: \mathcal{P}^L_2 \to \mathcal{P}^L_1$ which takes the $j$-th 
column $K^2_j$
of $\mathcal{P}^L_2$ to the $(j+1)$-th colume $K^1_{j+1}$ of $\mathcal{P}^L_1$
(both with an appropriate retraction); however, in Appendix A we prove the
following:
\begin{theo}
For every pair of mappings $\psi : \mathcal{P}^{L}_{1} \to 
\mathcal{P}^{L}_{2}$,
$\zeta :\mathcal{P}^{L}_{2} \to \mathcal{P}^{L}_{1}$ which make
$\mathcal{P}^{L}_{1}$ and $\mathcal{P}^{L}_{2}$ $k$-close, with $k < 
L/4$, there
exists a $p \in \mathcal{P}^{L}_{2}$ such that
$$
     D(p , \psi \circ \zeta (p) ) = L.
$$ \hfill$\square$
\end{theo}
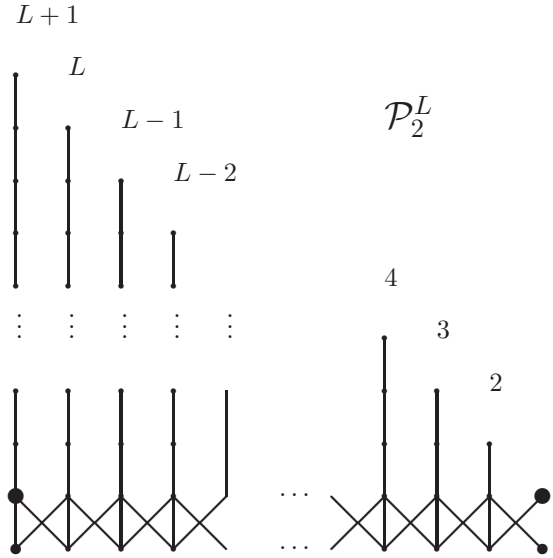
\begin{figure}[h]
\begin{center}
   \setlength{\unitlength}{0.7cm}
\begin{picture}(10,10)
\thicklines
\put(0,0){\line(0,1){1}}
\put(0,1){\line(1,-1){1}}
\put(1,0){\line(0,1){1}}
\put(1,1){\line(-1,-1){1}}
\put(1,1){\line(1,-1){1}}
\put(2,0){\line(0,1){}}
\put(2,1){\line(-1,-1){1}}
\put(2,1){\line(1,-1){1}}
\put(3,0){\line(0,1){1}}
\put(3,1){\line(-1,-1){1}}
\put(3,1){\line(1,-1){1}}
\put(3,0){\line(1,1){1}}
\put(5,0){\ldots}
\put(5,1){\ldots}
\put(7,0){\line(-1,1){1}}
\put(7,1){\line(-1,-1){1}}
\put(7,0){\line(1,1){1}}
\put(7,1){\line(1,-1){1}}
\put(8,0){\line(0,1){1}}
\put(8,0){\line(1,1){1}}
\put(8,1){\line(1,-1){1}}
\put(7,0){\line(0,1){1}}
\put(8,0){\line(0,1){1}}
\put(9,0){\line(0,1){1}}
\put(0,1){\line(0,1){1}}
\put(0,2){\line(0,1){1}}
\put(0,4){\vdots}
\put(0,5){\line(0,1){1}}
\put(0,6){\line(0,1){1}}
\put(0,7){\line(0,1){1}}
\put(0,8){\line(0,1){1}}
\put(1,1){\line(0,1){1}}
\put(1,2){\line(0,1){1}}
\put(1,4){\vdots}
\put(1,5){\line(0,1){1}}
\put(1,6){\line(0,1){1}}
\put(1,7){\line(0,1){1}}
\put(2,1){\line(0,1){1}}
\put(2,2){\line(0,1){1}}
\put(2,4){\vdots}
\put(2,5){\line(0,1){1}}
\put(2,6){\line(0,1){1}}
\put(3,1){\line(0,1){1}}
\put(3,2){\line(0,1){1}}
\put(3,4){\vdots}
\put(3,5){\line(0,1){1}}
\put(4,1){\line(0,1){1}}
\put(4,2){\line(0,1){1}}
\put(7,1){\line(0,1){1}}
\put(7,2){\line(0,1){1}}
\put(7,3){\line(0,1){1}}
\put(8,1){\line(0,1){1}}
\put(8,2){\line(0,1){1}}
\put(9,1){\line(0,1){1}}
\put(4,4){\vdots}
\put(0,0){\circle*{0.2}}
\put(0,1){\circle*{0.3}}
\put(1,1){\circle*{0.1}}
\put(1,0){\circle*{0.1}}
\put(2,0){\circle*{0.1}}
\put(2,1){\circle*{0.1}}
\put(3,0){\circle*{0.1}}
\put(3,1){\circle*{0.1}}
\put(7,0){\circle*{0.1}}
\put(7,1){\circle*{0.1}}
\put(8,0){\circle*{0.1}}
\put(8,1){\circle*{0.1}}
\put(9,0){\circle*{0.1}}
\put(9,1){\circle*{0.1}}
\put(0,2){\circle*{0.1}}
\put(0,3){\circle*{0.1}}
\put(0,5){\circle*{0.1}}
\put(0,6){\circle*{0.1}}
\put(0,7){\circle*{0.1}}
\put(0,8){\circle*{0.1}}
\put(0,9){\circle*{0.1}}
\put(1,2){\circle*{0.1}}
\put(1,3){\circle*{0.1}}
\put(1,5){\circle*{0.1}}
\put(1,6){\circle*{0.1}}
\put(1,7){\circle*{0.1}}
\put(1,8){\circle*{0.1}}
\put(2,2){\circle*{0.1}}
\put(2,3){\circle*{0.1}}
\put(2,5){\circle*{0.1}}
\put(2,6){\circle*{0.1}}
\put(2,7){\circle*{0.1}}
\put(3,2){\circle*{0.1}}
\put(3,3){\circle*{0.1}}
\put(3,5){\circle*{0.1}}
\put(3,6){\circle*{0.1}}
\put(7,2){\circle*{0.1}}
\put(7,3){\circle*{0.1}}
\put(7,4){\circle*{0.1}}
\put(8,2){\circle*{0.1}}
\put(8,3){\circle*{0.1}}
\put(9,2){\circle*{0.1}}
\put(9,1){\line(1,-1){1}}
\put(9,0){\line(1,1){1}}
\put(10,1){\circle*{0.3}}
\put(10,0){\circle*{0.2}}
\put(7,8){{\Large $ \mathcal{P}_2^{L}$}}
\put(0,10){$L+1$}
\put(1,9){$L$}
\put(2,8){$L-1$}
\put(3,7){$L-2$}
\put(7,5){$4$}
\put(8,4){$3$}
\put(9,3){$2$}
\end{picture}
\caption{Example 3}
\label{figure3}
\end{center}

\end{figure}
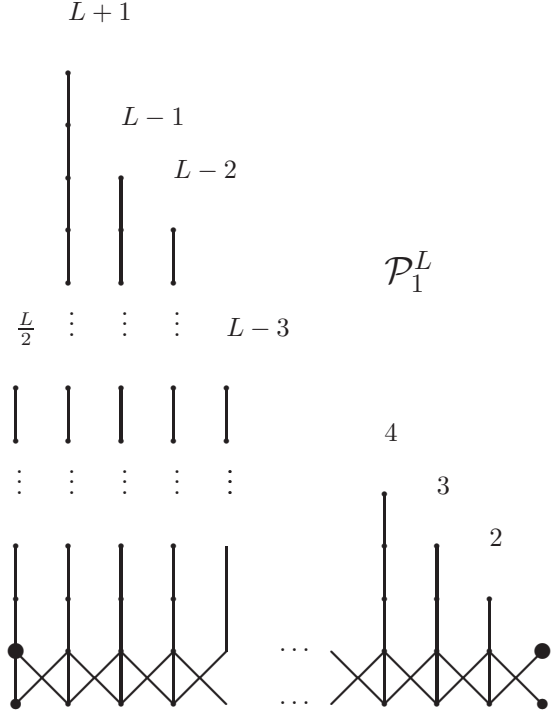
\begin{figure}[h]
\begin{center}
   \setlength{\unitlength}{0.7cm}
\begin{picture}(10,10)
\thicklines
\put(0,0){\line(0,1){1}}
\put(0,1){\line(1,-1){1}}
\put(1,0){\line(0,1){1}}
\put(1,1){\line(-1,-1){1}}
\put(1,1){\line(1,-1){1}}
\put(2,0){\line(0,1){}}
\put(2,1){\line(-1,-1){1}}
\put(2,1){\line(1,-1){1}}
\put(3,0){\line(0,1){1}}
\put(3,1){\line(-1,-1){1}}
\put(3,1){\line(1,-1){1}}
\put(3,0){\line(1,1){1}}
\put(5,0){\ldots}
\put(5,1){\ldots}
\put(7,0){\line(-1,1){1}}
\put(7,1){\line(-1,-1){1}}
\put(7,0){\line(1,1){1}}
\put(7,1){\line(1,-1){1}}
\put(8,0){\line(0,1){1}}
\put(8,0){\line(1,1){1}}
\put(8,1){\line(1,-1){1}}
\put(7,0){\line(0,1){1}}
\put(8,0){\line(0,1){1}}
\put(9,0){\line(0,1){1}}
\put(0,1){\line(0,1){1}}
\put(0,2){\line(0,1){1}}
\put(0,4){\vdots}
\put(0,5){\line(0,1){1}}
\put(1,1){\line(0,1){1}}
\put(1,2){\line(0,1){1}}
\put(1,4){\vdots}
\put(1,5){\line(0,1){1}}
\put(1,7){\vdots}
\put(1,8){\line(0,1){1}}
\put(1,9){\line(0,1){1}}
\put(1,10){\line(0,1){1}}
\put(1,11){\line(0,1){1}}
\put(2,1){\line(0,1){1}}
\put(2,2){\line(0,1){1}}
\put(2,4){\vdots}
\put(2,5){\line(0,1){1}}
\put(2,7){\vdots}
\put(2,8){\line(0,1){1}}
\put(2,9){\line(0,1){1}}
\put(3,1){\line(0,1){1}}
\put(3,2){\line(0,1){1}}
\put(3,4){\vdots}
\put(3,5){\line(0,1){1}}
\put(3,7){\vdots}
\put(3,8){\line(0,1){1}}
\put(4,1){\line(0,1){1}}
\put(4,2){\line(0,1){1}}
\put(4,4){\vdots}
\put(7,1){\line(0,1){1}}
\put(7,2){\line(0,1){1}}
\put(7,3){\line(0,1){1}}
\put(8,1){\line(0,1){1}}
\put(8,2){\line(0,1){1}}
\put(9,1){\line(0,1){1}}
\put(4,4){\vdots}
\put(0,0){\circle*{0.2}}
\put(0,1){\circle*{0.3}}
\put(1,1){\circle*{0.1}}
\put(1,0){\circle*{0.1}}
\put(2,0){\circle*{0.1}}
\put(2,1){\circle*{0.1}}
\put(3,0){\circle*{0.1}}
\put(3,1){\circle*{0.1}}
\put(7,0){\circle*{0.1}}
\put(7,1){\circle*{0.1}}
\put(8,0){\circle*{0.1}}
\put(8,1){\circle*{0.1}}
\put(9,0){\circle*{0.1}}
\put(9,1){\circle*{0.1}}
\put(0,2){\circle*{0.1}}
\put(0,3){\circle*{0.1}}
\put(0,5){\circle*{0.1}}
\put(0,6){\circle*{0.1}}
\put(1,2){\circle*{0.1}}
\put(1,3){\circle*{0.1}}
\put(1,5){\circle*{0.1}}
\put(1,6){\circle*{0.1}}
\put(1,8){\circle*{0.1}}
\put(1,9){\circle*{0.1}}
\put(1,10){\circle*{0.1}}
\put(1,11){\circle*{0.1}}
\put(1,12){\circle*{0.1}}
\put(2,2){\circle*{0.1}}
\put(2,3){\circle*{0.1}}
\put(2,5){\circle*{0.1}}
\put(2,6){\circle*{0.1}}
\put(2,8){\circle*{0.1}}
\put(2,9){\circle*{0.1}}
\put(2,10){\circle*{0.1}}
\put(3,2){\circle*{0.1}}
\put(3,3){\circle*{0.1}}
\put(3,5){\circle*{0.1}}
\put(3,6){\circle*{0.1}}
\put(7,2){\circle*{0.1}}
\put(7,3){\circle*{0.1}}
\put(7,4){\circle*{0.1}}
\put(8,2){\circle*{0.1}}
\put(8,3){\circle*{0.1}}
\put(9,2){\circle*{0.1}}
\put(9,1){\line(1,-1){1}}
\put(9,0){\line(1,1){1}}
\put(10,1){\circle*{0.3}}
\put(10,0){\circle*{0.2}}
\put(3,8){\circle*{0.1}}
\put(3,9){\circle*{0.1}}
\put(4,5){\line(0,1){1}}
\put(4,5){\circle*{0.1}}
\put(4,6){\circle*{0.1}}
\put(4,7){$L-3$}
\put(7,8){{\Large $ \mathcal{P}_1^{L}$}}
\put(0,7){$\frac{L}{2}$}
\put(1,13){$L+1$}
\put(2,11){$L-1$}
\put(3,10){$L-2$}
\put(7,5){$4$}
\put(8,4){$3$}
\put(9,3){$2$}
\end{picture}
\caption{Example 3}
\label{figure4}
\end{center}
\end{figure}

Now choose $\alpha > 0$ and let $\epsilon_{m} = \frac{4\alpha}{4m + 1}$, $L_{m}
= 4m$, $m \in \NN \setminus \left\{ 0,1 \right\}$.  Define for $i=1,2$ the
discrete Lorentz spaces $\mathcal{P}_{i}^{\epsilon_{m}, L_{m}}$ by the same
Hasse diagrams, but now suppose that every link has length $\epsilon_{m}$.
Arguments analogous to the ones around theorem 2 tell us that
$\mathcal{P}_1^{\epsilon_m,L_m}$ and $\mathcal{P}_2^{\epsilon_m,L_m}$
are $\epsilon_{m}$-close, but for every pair of mappings $\psi_m :
\mathcal{P}^{\epsilon_{m},L_{m}}_1 \to \mathcal{P}^{\epsilon_m,L_m}_2$,
$\zeta_m :\mathcal{P}^{\epsilon_{m},L_{m}}_{2} \rightarrow
\mathcal{P}^{\epsilon_{m},L_{m}}_{1}$ which make
$\mathcal{P}^{\epsilon_{M},L_{m}}_{1}, \mathcal{P}^{\epsilon_{m},L_{m}}_{2}$
$\epsilon$- close, with $\epsilon < \frac{4m\alpha}{4m+1}$, there exists a $p
\in \mathcal{P}^{\epsilon_m,L_m}_2$ such that
$$
     D_{m}(p , \psi_{m} \circ \zeta_{m} (p) ) = \frac{16m\alpha}{4m+1}\;.
$$
However, $tdiam(\mathcal{P}_{i}^{\epsilon_{m}, L_{m}}) = 4 \alpha$ for $i=1,2$
and $m >1$.\hfill$\square$
\\*
\\*
In other words, $\mathcal{P}_1^{\epsilon_m,L_m}$ and
$\mathcal{P}_2^{\epsilon_m,L_m}$ are $\epsilon$-close with $\epsilon \to 0$ as
$m \to \infty$, but for every pair of maps $\psi$ and $\zeta$ that realize the
closeness condition, there exists a $p \in \mathcal{P}^{\epsilon_m,L_m}_2$ such
that $D_m(p, \psi_m\circ \zeta_m(p)) \to 4\alpha$. This proves our 
claim that the
qualitative uniformities defined by GH and GGH are inequivalent.

However, this does not prove yet that there exist GH Cauchy sequences which
are not GGH.  In fact, we show now that if a Lorentz space $(\mathcal{M},d)$
is a limit space of a GH Cauchy sequence $\{(\mathcal{M}_i,d_i)\}_{i \in
\NN}$, then this sequence is GGH Cauchy and converges to the same limit (up to
isometry).  An intermediate result is the following.
\begin{theo}
Any isometry $\psi$ on a Lorentz space $(\mathcal{M},d)$ is a bijection.
\end{theo}
\textsl{Proof:}
\\*
We only have to show that $\psi$ is a surjection since, obviously, it is an
injection.  Evidently, $D(\psi(p), \psi(q)) \geq D(p,q)$ for all $p,q \in
\mathcal{M}$.  Suppose we can find an open ball $B_{D}(r, \epsilon)$ which is
not in $\psi(\mathcal{M})$; then $\psi^{k}(r) \notin B_{D}(\psi^{l}(r),
\epsilon)$ for all $k > l$.  Since $\mathcal{M}$ is compact, we may, by
passing to a subsequence if necessary, assume that $\psi^{l}(r) \stackrel{l
\rightarrow \infty}{\rightarrow} \psi^{\infty}(r)$.  Hence, we arrive at the
contradiction that $\psi^{\infty}(r) \notin B_{D}(\psi^{\infty}(r),
\epsilon)$.  This implies that all isolated points belong to
$\psi(\mathcal{M})$.  But then we have that
$$
     D(\psi(p), \psi(q)) = \sup_{r \in \mathcal{M}}
     \left| d(\psi(r),\psi(p)) + d(\psi(p),\psi(r)) - d(\psi(r),
     \psi(q)) - d(\psi(q) , \psi(r)) \right| .
$$
The rhs of this equation equals $D(p,q)$.  This shows that $\psi$ is surjective
since all isometries of compact metric spaces are. \hfill$\square$
\\*
\\*
Before we prove the main result we still need the following.
\begin{theo}
Let $\left\{\psi_{i} \mid i\in\NN_{0} \right\}$ be a set of
$\frac{1}{i}$-isometries on $\mathcal{M}$.  Then there exists a subsequence
$\{\psi_{i_n}\}_{n \in \NN}$ which uniformly converges in the strong sense to
an isometry $\psi$.
\end{theo}
\textsl{Proof:}
\\*
As usual, let $\mathcal{C}$ be a countable dense subset of $\mathcal{M}$ and
let $\{\psi_{i_n}\}_{n\in\NN}$ be a subsequence such that $\psi_{i_n}(p)
\stackrel{n \rightarrow \infty}{\rightarrow} \psi(p)$ for all $p \in
\mathcal{C}$.  It is easy to see that $\psi$ has a unique extension to a
$D$-isometry (and $d$ isometry) using Theorem 3.  The proof of Theorem 3 also
implies that $\psi(\mathcal{C})$ is dense in $\mathcal{M}$.   As a
consequence, we have that for any $\epsilon > 0$ there exists a $k(\epsilon) >
0$ such that $\psi_{i_{k}} ( \mathcal{C} )$ is $\epsilon$-dense in
$\mathcal{M}$ for $k > k(\epsilon)$.   Hence, $\left| D(\psi_{i_{k}}(p),
\psi_{i_{k}}(q)) - D(p,q) \right| < \epsilon + \frac{2}{i_{k}}$ for $k >
k(\epsilon)$ and for all $p,q \in \mathcal{M}$.  This implies that
$$
     D(\psi(r), \psi_{i_{k}}(r)) \leq
     D(\psi(p) , \psi_{i_{k}}(p)) + 2D(p,r) + \frac{2}{i_{k}} + \epsilon\;,
$$
for $k > k(\epsilon)$ and $p \in \mathcal{C}$. Since $\epsilon$ and $p$ can be
independently chosen arbitrarily close to $0$ and $r$ respectively, the result
follows. \hfill$\square$ \\*
\\*
We are now in a position to prove the main result.
\begin{theo}
Let $\{(\mathcal{M}_i,d_i)\}_{i\in\NN}$ be a GH Cauchy sequence of Lorentz
spaces converging to a Lorentz space $(\mathcal{M},d)$; then this sequence is
GGH Cauchy and converges to the same limit space.
\end{theo}
\textsl{Proof:}
\\*
Choose $\delta > 0$; then Theorem 4 implies that there exists a $\gamma >0$,
such that if $f$ is a $\gamma$-isometry, then there exists an isometry $g$
such that
$$
     D(f(x),g(x)) < {\textstyle\frac{\delta}{2}} \qquad \forall x \in \man.
$$
Let $\psi_{i} : \mathcal{M}_{i} \rightarrow \mathcal{M}$ and $\zeta_{i} :
\mathcal{M} \rightarrow \mathcal{M}_{i}$ be mappings which make
$(\mathcal{M}_{i},d_{i})$ and $(\mathcal{M},d)$ $\epsilon_{i}$-close, where
$\epsilon_{i} \stackrel{i \rightarrow \infty}{\rightarrow} 0$.  Then the
previous remark implies that for $i$ large enough that $2
\epsilon_{i} < \min \left\{\gamma, \frac{\delta}{2} \right\}$, there exists an
isometry $\beta_{i}$ such that
$$
     D(\beta_{i}(x), \psi_{i} \circ \zeta_{i} (x))
     < {\textstyle\frac{\delta}{2}} \qquad \forall x \in \man\;,
$$
or
$$
     D(x, \psi_{i} \circ \zeta_{i} \circ \beta_{i}^{-1}(x))
     < {\textstyle\frac{\delta}{2}} \qquad \forall x \in \man\;.
$$
Hence
$$
     D_{i}(p_{i},\zeta_{i} \circ \beta_{i}^{-1} \circ \psi_{i} (p_{i}))
     \leq 2 \epsilon_{i} + D(\psi_{i}(p_{i}), \psi_{i} \circ \zeta_{i} \circ
     \beta_{i}^{-1} \circ \psi_{i} ( p_{i}))\;,
$$
which implies that
$$
     D_{i}(p_{i} , \zeta_{i} \circ \beta_{i}^{-1} \circ \psi_{i} ( p_{i}))
     \leq 2 \epsilon_{i} + {\textstyle\frac{\delta}{2}} < \delta\;.
$$
Hence, for $i$ sufficiently large, $\psi_{i}$ and $\zeta_{i} \circ
\beta_{i}^{-1}$ make $(\mathcal{M}_{i},d_{i})$ and $(\mathcal{M},d)$,
$(\epsilon_{i}, \delta)$-close.  \hfill$\square$ \\*
\\*
This does not show that every GH Cauchy sequence is GGH, but one which is not
GGH either has no sensible limit, or its limit is not ``spatially compact".
This last theorem also has as a consequence that the trivial extension of
$d_{\rm GH}$ to the moduli space of isometry classes of Lorentz spaces is a
metric.

\section{Some properties of the strong metric}

In this section, we study some properties of the strong metric (\ref{D})
introduced in Ref \cite{Noldus1}, with the aim of understanding better its
relationship with the Lorentz distance. It was remarked in Ref \cite{Noldus1}
that the topologies induced by these two distances are equivalent on globally
hyperbolic spacetimes. A natural question to ask therefore is whether in fact
the distances are equivalent, in the sense that the strong metric determines
the Lorentz metric, up to time reversal (obviously, applying a ``time reversal"
$d'(x,y) = d(y,x)$ to a Lorentz distance produces a different Lorentz
distance with the same associated strong metric). We shall only begin to deal
with this question here, by giving an example which shows that the answer in
general is no, but as a beginning in the study of this relationship, most of
this section will be devoted to the simpler question, What is the shape of open
ball $B_D(p,\epsilon)$ of radius $\epsilon$ around a point $p$ in the strong
metric?"
\\*
\\*
To start with, we ``split" the strong metric $D$ into two
pseudodistances $D^\pm$ which will be useful later on, by defining
$$
        D^+( p,q) = \max_{r \in \mathcal{M}}
        \big| d_{g}(p,r) - d_{g}(q,r) \big|
$$
and
$$
        D^-( p,q)
        = \max_{r \in \mathcal{M}} \big| d_{g}(r,p) - d_{g}(r,q) \big|\;,
$$
in terms of which $D$ can be recovered as \cite{Noldus2}
$$
        D(p,q) = \max \left\{ D^+(p,q) , D^-(p,q) \right\}.
$$
Notice that, although individually $D^+$ and $D^-$ are \emph{pseudo} metrics
($D^{\pm}(p,q) = 0$ does not necessarily imply that $p=q$), this limitation
of $D^+$ ($D^-$) arises only for $p$ and $q$ both belonging to the 
future (past)
boundary of $\man$.  For example, clearly $D^+(p,q) = 0$ for all $p,q \in
\partial_{\rm F}\man$, but if $p \not\in \partial_{\rm F}\man$ and $q\in
\partial_{\rm F}\man$, any $r \in I^+(p)$ gives $d_g(p,r) = |d_g(p,r)-d_g(q,r)|
>  0$, and if both $p,q\not\in \partial_{\rm F}\man$, the same is true with any
$r \in I^+(p) \setminus I^+(q)$ (assuming $p \not\in J^+(q)$, otherwise just
switch $p$ and $q$).
\\* \\*
These remarks show that both $D^\pm$ are true distances on the interior of
$\man$, and they also motivate us to try to locate the ``distance-maximizing
points", i.e., the points $r$ which realize the maximum in the 
definition of those
functions for given $p$ and $q$.  Such points will then allow us to control
the non-locality in $D^\pm$ and $D$, which is what makes the study of their
detailed properties more difficult than those of a distance arising from
a positive-definite metric tensor.
\\*

\noindent\textbf{Property} {\em Given any two points $p$ and $q$ not both
belonging to $\partial_{\rm F} \mathcal{M}$, any point $r$ such that $D^+(p,q)
= |d_{g}( p,r) -  d_{g}(q,r)|$ is an element of $I^{+} (p) \,\triangle\,
I^{+}(q)$,\footnote{For any two sets $A$ and $B$, $A\,\triangle\,B$ stands for
the symmetric difference $(A \setminus B) \cup (B \setminus A)$.} and $I^+(r)
\subset I^+(p) \cap I^+(q)$.  A dual property holds for $D^-$.}
\\*
\\*
\noindent\textsl{Proof:}
\\*
\noindent Obviously, $r \in I^+(p) \cup I^+(q)$. Suppose $r \in I^{+}(p) \cap
I^{+}(q)$ and, without loss of generality, assume that $d_{g}(p,r) >
d_{g}(q,r)$.  Let $\gamma$ be a distance-maximizing geodesic from $p$ to
$r$; then $\gamma$ cuts $E^{+}(q)$ in a point $s$.  But then, the reverse
triangle inequality implies that
\begin{eqnarray*}
        d_g(p,r) - d_g(q,r) &=& d_g(p,s) + d_g(s,r) - d_g(q,r) \\
        &<& d_g(p,s)\ \ =\ \ d_g(p,s) - d_g(q,s)\;,
\end{eqnarray*}
so $r$ cannot be distance-maximizing.  The second statement says that $r$
belongs to the boundary of the light cone of $p$ or $q$.  Again, without loss
of generality assume that $r \in I^+(p) \setminus I^+(q)$ (in this case, $p
\not\in \partial_{\rm F} \man$).  Then if the statement is false we can extend
the distance-maximizing geodesic from $p$ to $r$ to a new point $r'$ such that
$d_g(p,r')>d_g(p,r)$ but still $d_g(q,r') = 0$, contrary to the assumption.
\hfill$\square$
\\*
\\*
Now, if $(\mathcal{M},g)$ contains no cut points, then $r$ must belong to
$\partial_{\rm F} \mathcal{M}$.  Since suppose $r \in I^{+}(p) \setminus
I^{+}(q)$, but does not belong to $\partial_{\rm F}\man$; then $I^{+}(r)
\subset I^{+}(p) \cap I^{+} (q)$ implies that $r$ belongs to $E^{+}(q)$.
Let $\gamma$ be the unique null geodesic from $q$ to $r$, then moving $r$ to
the future along this null geodesic up to $\partial_{\rm F} \mathcal{M}$ keeps
$r$ out of $I^{+}(q)$, otherwhise the geodesic would have a cut point, which
is contrary to the assumption.  The following theorem is also valid when there
are cut  points, but then the statement can be made sharper.
\\*
\\*
As was remarked earlier \cite{Noldus1}, the $\epsilon$-balls $B_D(p,\epsilon)$
are causally convex, in the sense that if $x,y \in B_D(p,\epsilon)$, then
$A(x,y) \subset B_D(p,\epsilon)$.  We now wish to find out more about those
sets. To begin with, notice that
$$
       B_D(p,\epsilon) = B_{D^+}(p,\epsilon) \cap B_{D^-}(p,\epsilon)\;.
$$
Then we have:

\begin{theo}

Let $(\mathcal{M},g)$ be a spacetime with no cut points and choose a point\\*
$p \in \mathcal{M} \setminus \partial_{\rm F} \mathcal{M}$ and an $\epsilon >
0$ such that $K^{+} ( p , \epsilon ) \neq \emptyset$.  Then the open ``sphere"
$B_{D^+} (p,\epsilon)$ of radius $\epsilon$ centred at $p$ with respect to
the  pseudometric $D^{+}$ satisfies
$$
       B_{D^+} (p , \epsilon) \subseteq \left[ \bigcap_{x \in \mathcal{H}^{+}
       (p)}  \left( \mathcal{O}^{-} (x, \epsilon ) \right)^{\rm c} \right]
       \bigcap \left[
       \bigcap_{x \in \mathcal{F}^{+} (p , \epsilon )} I^{-} (x) \right],
$$
where
\begin{itemize}

\item $K^{+} (x , \epsilon) = \left\{ y \in \mathcal{M} \mid d_{g}
(x,y) = \epsilon \right\}$, the future $\epsilon$-sphere centred at $x$,

\item $\mathcal{O}^{-} (x , \epsilon) = \left\{ y \in \mathcal{M}
\mid d_{g} (y,x) \geq \epsilon \right\}$, the closed outer past
$\epsilon$-ball around $x$,

\item $\mathcal{H}^{+}(p) = E^{+}(p) \cap \partial_{\rm F} \mathcal{M}$,

\item $\mathcal{F}^{+}(p , \epsilon) = K^{+}( p ,\epsilon) \cap
\partial_{\rm F} \mathcal{M}$.

\end{itemize}

\noindent The open sphere $B_{D^-} ( p, \epsilon )$ defined with respect to
the pseudometric $D^{-}$ satisfies a similar inclusion property with all
pasts and futures interchanged.
\end{theo}

\noindent\textsl{Proof:}
\\*
\noindent Let $x \in B_{D^+} (p , \epsilon)$; then $x$ must be chronologically
connected to all points in $\mathcal{F}^{+}(p , \epsilon)$.  For, suppose there
exists a point $y \in \mathcal{F}^{+}(p , \epsilon)$ such that $x \notin
I^{-}(y)$; then $d_{g}(p,y) - d_{g}(x,y)  = \epsilon$, which is a 
contradiction.
On the other hand, $x$ cannot belong to $\mathcal{O}^-(z,\epsilon)$ for any $z
\in \mathcal{H}^+(p)$, since otherwise
$$
        d_{g} ( x , z ) - d_{g}(p,z) \geq \epsilon\;,
$$
which is impossible. \hfill$\square$
\\*

\noindent The following figure 2 shows that the above inclusion can be an
equality.  The universe is $(S^{1} \times \left[ 0 , 1\right], -\dd t^{2} +
\dd\theta^{2})$  and the shaded area represents $B_{D^+} (p, \epsilon)$ for
$\epsilon$ sufficiently small.

\begin{figure}[h]
	\begin{center}
			\scalebox{0.50}{\includegraphics{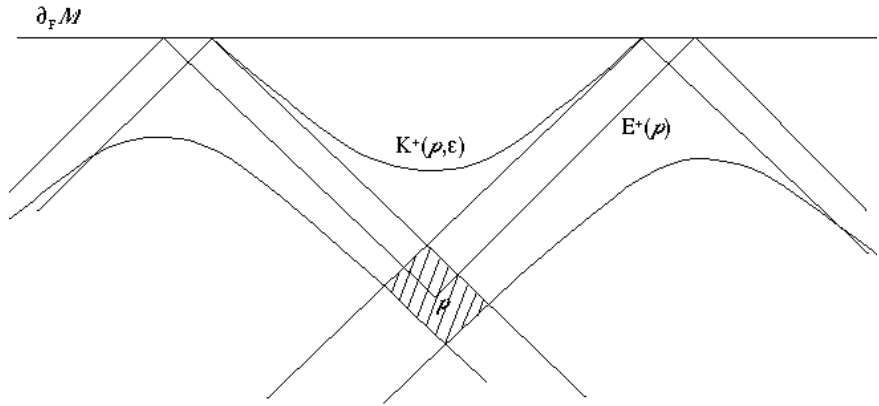}}
   	\end{center}
\caption{Illustration of theorem 2}
\label{fig:grliph5-3}

\end{figure}
Obviously, a dual statement holds for $B_{D^-}(p,\epsilon)$. Putting together
these two bounds on $B_{D^\pm}$, we also notice that for $B_D$ we can 
write down
a simpler, but weaker bound
$$
       B_D(p,\epsilon) \subseteq
       \bigcap_{x\in\mathcal{F}^-(p,\epsilon),\,y\in\mathcal{F}^+(p,\epsilon)}
       A(x,y)\;,
$$
which says that $B_D(p,\epsilon)$ is contained in all of the ``long skinny"
(for small $\epsilon$) Alexandrov neighborhoods defined by pairs of points
on the past and future boundaries of $M$ which are ``almost null 
related to $p$".
\\*
\\*
Concerning the second question posed at the beginning of this section, we
notice that the strong metric $D$ does not determine the Lorentz distance $d$
up to time reversal for discrete Lorentz spaces.
\begin{example}
\end{example}
Consider the causal sets $\mathcal{P}_{1}$ and $\mathcal{P}_{2}$ defined by
the Hasse diagrams below.
\begin{figure}[h]
\begin{center}
   \setlength{\unitlength}{0.7cm}
\begin{picture}(3.5,2)
\thicklines
\put(0,0){\line(0,1){1}}
\put(0,1){\line(1,-1){1}}
\put(1,0){\line(0,1){1}}

\put(0,0){\circle*{0.1}}
\put(0,1){\circle*{0.1}}
\put(1,1){\circle*{0.1}}
\put(1,0){\circle*{0.1}}
\put(2,0){\circle*{0.1}}

\put(1, 1.5){{\Large $ \mathcal{P}_1$}}
\end{picture}
\begin{picture}(2,2)
\thicklines
\put(0,0){\line(0,1){1}}
\put(0,1){\line(1,-1){1}}
\put(1,0){\line(0,1){1}}
\put(1,1){\line(1,-1){1}}

\put(0,0){\circle*{0.1}}
\put(0,1){\circle*{0.1}}
\put(1,1){\circle*{0.1}}
\put(1,0){\circle*{0.1}}
\put(2,0){\circle*{0.1}}

\put(1, 1.5){{\Large $\mathcal{P}_2$}}
\end{picture}
\end{center}
\caption{Hasse diagrams}
\label{fig7}
\end{figure}
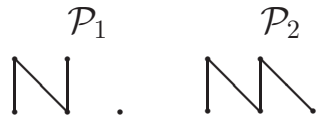
Clearly $d_{GH}((\mathcal{P}_{1},d_{1}),(\mathcal{P}_{2},d_{2})) = 1$, while
$d_{GH}((\mathcal{P}_{1},D_{1}),(\mathcal{P}_{2},D_{2})) = 0$ since in both
posets we have $D_i(p,q) = 1$ for all distinct $p$ and $q$.  Moreover,
$D_1^+ \neq D_2^+$ while $D_{1}^{-} = D_{2}^{-}$.  This clearly shows
that convergence in the strong metric is not sufficient to guarantee
convergence of the Lorentz metrics at least as far as Lorentz spaces are
concerned.\hfill$\square$

\section{Causal properties of the limit space}

In this section, we study further causal properties of the limit spaces of
sequences of Lorentzian globally hyperbolic spacetimes that were introduced
in Ref \cite{Noldus2}. The most daunting problem is how to define a causal
relation on the degenerate regions.  First of all, one might ask whether
this is a physically meaningful question, in the sense of whether any
relation we may define can give rise to sensible causal curves along which
``particles" travel, given the fact that this causal structure may be
very non-local and depend on what the limit space is like both to the future
and to the past of the degenerate region.
We cannot stress the word \emph{global} enough; for example, even a
small change in the future lightcones in a ``distant" region according to
one set of observers could be spotted by the strong metric in a ``tiny"
Alexandrov neighborhood which is ``boost equivalent" to a neighborhood
that is more localized for a set of observers used to realize the bound
in the definition of the strong distance.\footnote{Notice that if $p$ is
a point of the past boundary, then $x \rightarrow D^{-}(p,x)$ is a 
time function.}
Hence, it is not entirely clear whether defining a causal relation is 
physically
meaningful or not. However, it is for sure a quite interesting 
mathematical problem
and it shall be treated as such in the rest of this section.
\\*
\\*
In the sequel we will call a proposal for the causal relation \emph{good} if,
roughly speaking, it coincides on limit spaces of arbitrary GGH Cauchy
sequences of conformally equivalent spacetimes with the causal relations
defined by the elements of these sequences.  To start with, we give an example
which shows that the closure of the space of discrete Lorentz spaces contains
spaces whose causal behaviour differs significantly from the kind of limit
spaces already considered before \cite{Noldus2}.\footnote{A similar kind of
limit space was communicated to the second author by Rafael Sorkin.}
\begin{example}
\end{example}
Suppose that $L > 2$ and let $(\mathcal{P}^{L}_{n},d_{n})$ be a 
discrete Lorentz
space defined by the elements $\left\{ \frac{iL}{n} \mid i=0 \ldots n \right\}$
and $d_{n}(p,q) = \max \{ 0 , q - p - 1 \}$. It is easy to check
that $d_{n}$ defines a Lorentz distance.  Moreover
$(\mathcal{P}^{L}_{n},d_{n}) \stackrel{n \rightarrow \infty}{\rightarrow}
(\left[0,L \right], d)$ where, evidently, $d(p,q) = \max \left\{ 0 , q-p-1
\right\}$.  Obviously, we want the causal relation to be the ordinary order
relation on $\left[0,L\right]$.  Hence, any pair of timelike related points in
the limit space can \emph{only} be connected by a causal curve $\gamma$ which
is nowhere timelike in the sense that for any $t$ on $\gamma$, $0 < s - t < 1$
implies that $d(t,s) = 0$.  The strong metric $D(t,s)$ between points $t < s$
equals $s - t$ unless $0 < t < 1$ and $L-1 < s < L$, in which case it equals
$\max\{L-t,s\} - 1$. Hence, locally, $D$ is the path metric defined
by the standard line element $\dd t^{2}$.  Conclusion: although the limit
space has a manifold structure, the Lorentzian distance is far from being
derivable from a tensor. \hfill$\square$
\\*
\\*
In this example, we defined the causal relation using our intuition.  Since we
are looking for a general prescription, we might postulate something like $p
\leq q$ iff $I^{+}(q) \subset I^{+}(p)$ and $I^{-}(p) \subset I^{-}(q)$.
However, as mentioned in Ref \cite{Noldus2}, this is not sufficient.
Therefore, let us start by defining the causal relation on the subset of
$\mathcal{M}$ that we are most familiar with.  As announced at the end of
\cite{Noldus2}, a good candidate for $\leq$ on the closure
$\overline{\mathcal{TCON}}$ of the timelike continuum is the $K^{+}$ causal
relation defined by Sorkin and Woolgar \cite{SW}, i.e.,

\begin{deffie}
$K^{+}$ is the smallest topologically closed partial order in $\man 
\times \man$
containing $I$.
\end{deffie}

We need to remark here that, as mentioned in \cite{SW}, $K^+$ can be 
defined as the
relationship $\prec$ built by transfinite induction with the 
following procedure:
\begin{itemize}
\item $\prec^0 = I^+$;
\item $\prec^\alpha = \bigcup_{\beta<\alpha}\prec^\beta$ if $\alpha$ is a limit
ordinal;
\item $\prec^{\beta+1}$ is constructed from $\prec^\beta$ by adding 
pairs which are
implied either by transitivity or by closure.
\end{itemize}

Since $\man \times \man$ has at most $2^{\aleph_0}$ elements, the 
procedure must
terminate at an ordinal\footnote{For more information about ordinals 
and transfinite
induction, see \cite{Kelley}.} with cardinality less than or equal to
$2^{\aleph_0}$. The following example illustrates that it can run up 
to an ordinal
of cardinality $\aleph_0$.

\begin{example}
\end{example}
Let ${\bf N} = \aleph_0$, $\mathcal{P} = \big\{w^{i},\, x^{j}_{k},\, 
y^{j}_{k},\,
z^{j} \mid i \in {\bf N}+1 \textrm{ and } j,k \in {\bf N} \big\}$, 
and construct
the discrete Lorentz space $(\mathcal{P},d)$ by defining
\begin{eqnarray}
     & &d(w^{i},z^{j}) = d(x^{i}_{k}, z^{j}) = \frac{1}{(j+1)^2}\;, \nonumber\\
     & &d(y^{i}_{k}, z^{j+1}) = \frac{1}{(j+2)^2}\;, \qquad
     d(x^{i}_{j},y^{i}_{j}) = \frac{1}{(j+1)(i+1)^{2}}\;,
\end{eqnarray}
for all $k$ and $i \leq j$ in ${\bf N}$; all other distances are 
calculated from
these values by taking the maximum over all ``timelike'' chains. From 
these data,
it is easy to calculate the strong distance:
\begin{eqnarray}
     & &D(w^{i},x^{i}_{j}) = D(w^{i+1},y^{i}_{j})
     = \frac{1}{(j+1)(i+1)^{2}} \nonumber\\
     & &D(x^{i}_{j},y^{i}_{j}) = \frac{1}{(i+1)^{2}}\;.
\end{eqnarray}
We now start our program: $\prec^{1}$ is the closure of $I^{+}$. 
Obviously, the new
relations induced by this procedure are $w^{i} \prec^{1} w^{i+1}$. 
$\prec^{2}$ is
constructed from $\prec^{1}$ by adding pairs which are implied by 
transitivity and
closure: this results in $w^{i} \prec^{2} w^{i+2}$ for all $i \in {\bf N}$ and
$w^{\bf N}  \prec^{2} w^{\bf N}$.  The reader may easily check that at stage
$n>2$, $\prec^{n}\ =\ \prec^{n-1} \cup \left\{(w^{i}, w^{j}) \mid i+2^{n-1} < j
\leq i + 2^{n} \in {\bf N} \right\}$.  Hence,
$$
    \prec^{\bf N} = I^{+} \cup \left\{(w^{i}, w^{j}) \mid i < j
    \in {\bf N} \right \} \cup \left\{(w^{\bf N}, w^{\bf N}) \right\}.
$$
But this relation is not closed yet, and
$$
    \prec^{{\bf N}+1} = I^{+} \cup \left\{(w^{i}, w^{j}) \mid i < j 
\in {\bf N}+1
    \right \} \cup \left\{(w^{\bf N}, w^{\bf N}) \right\}.
$$
So the procedure stops at the $({\bf N}+1)$-st step and the 
cardinality of ${\bf
N}+1$ is $\aleph_0$. \hfill$\square$
\\*
\\*
Since $K$ gives in general more information than $I$,\footnote {One
can construct Lorentz spaces where $K^{+}(q) \subseteq K^{+}(p)$ and
$K^{-}(p) \subseteq K^{-} (q)$ do \emph{not} imply that $I^{+}(q) \subseteq
I^{+}(p)$ and $I^{-}(p) \subseteq I^{-}(q)$ and vice versa. However, $K^{+}(q)
\subseteq K^{+}(p)$ and $K^{-}(p) \subseteq K^{-} (q)$ do imply that
$I^{+}(q) \subseteq I^{+}(p)$ and $I^{-}(p) \subseteq I^{-}(q)$ for Lorentz
spaces $(\mathcal{M},d)$ satisfying the following division property:
$$
     \forall p \ll q,\ \exists r :\ p \ll r \ll q\;.
$$
} we might hope that adding the conditions $K^{+}(q) \subset K^{+}(p)$ and
$K^{-}(p) \subset K^{-}(q)$ in order for $p \leq q$ would lead to a satisfying
definition.  Unfortunately it does not, as illustrated by the 
following example.
\begin{example}
\end{example}
To simplify the notation in this discussion, define a candidate relation
$\mathcal{R}$ between $p,\, q \in \man \setminus \overline{\mathcal{TCON}}$ as
follows:
\begin{eqnarray*}
     p\,\mathcal{R}\,q &\Leftrightarrow& K^{+}(q) \subseteq K^{+}(p)\;,
     \ K^{-}(p) \subseteq K^{-} (q)\;, \\
     & & I^{+}(q) \subseteq I^{+}(p) \textrm{ and }
     I^{-}(p) \subseteq I^{-}(q)\;.
\end{eqnarray*}
Figure 7 shows a Lorentz space which is part of the cylinder, with
degenerate regions indicated by the shaded areas. The points $p$ and $q$ are
such that $p\, \mathcal{R}\, q$, while clearly we do not want that $p \leq q$.
However, there exists no curve $\gamma$ between $p$ and $q$ satisfying the
condition that $\gamma(t)\, \mathcal{R}\, \gamma(s)$ for all $t \leq s$.
\hfill$\square$
\begin{figure}[h]
\begin{center}
  \setlength{\unitlength}{1.7cm}
\begin{picture}(8,3.2)

\put(1,1){\line(1,0){6}}
\put(1,1){\line(0,1){1.91}}
\put(1,2.91){\line(1,0){6}}
\put(7,1){\line(0,1){1.91}}
\put(1,0.5){$0$}
\put(7,0.5){$2 \pi$}
\put(3.53,1.96){\circle*{0.05}}
\put(4.45,1.96){\circle*{0.05}}
\thicklines
\put(3.53,1.96){\line(1,1){0.95}}
\put(3.53,1.96){\line(1,-1){0.95}}
\put(3.53,1.96){\line(-1,1){0.95}}
\put(3.53,1.96){\line(-1,-1){0.95}}
\put(4.45,1.96){\line(1,1){0.95}}
\put(4.45,1.96){\line(1,-1){0.95}}
\put(4.45,1.96){\line(-1,1){0.95}}
\put(4.45,1.96){\line(-1,-1){0.95}}

\thinlines
\put(3.53,2.06){\Blue{\line(1,-1){0.5}}}
\put(3.43,1.96){\Blue{\line(1,-1){0.5}}}
\put(3.38,1.91){\Blue{\line(1,-1){0.5}}}
\put(3.33,1.86){\Blue{\line(1,-1){0.5}}}
\put(3.28,1.81){\Blue{\line(1,-1){0.5}}}
\put(3.23,1.76){\Blue{\line(1,-1){0.5}}}
\put(3.18,1.71){\Blue{\line(1,-1){0.5}}}
\put(3.13,1.66){\Blue{\line(1,-1){0.5}}}
\put(3.08,1.61){\Blue{\line(1,-1){0.5}}}
\put(3.03,1.56){\Blue{\line(1,-1){0.5}}}
\put(2.98,1.51){\Blue{\line(1,-1){0.5}}}
\put(2.93,1.46){\Blue{\line(1,-1){0.45}}}
\put(2.88,1.41){\Blue{\line(1,-1){0.40}}}
\put(2.83,1.36){\Blue{\line(1,-1){0.35}}}
\put(2.78,1.31){\Blue{\line(1,-1){0.30}}}
\put(2.73,1.26){\Blue{\line(1,-1){0.25}}}
\put(2.68,1.21){\Blue{\line(1,-1){0.20}}}
\put(2.63,1.16){\Blue{\line(1,-1){0.15}}}
\put(2.58,1.11){\Blue{\line(1,-1){0.10}}}
\put(2.53,1.06){\Blue{\line(1,-1){0.10}}}
\put(2.48,1.01){\Blue{\line(1,-1){0.10}}}
\put(2.43,0.96){\Blue{\line(1,-1){0.10}}}

\put(3.93,2.36){\Blue{\line(1,-1){0.5}}}
\put(4.03,2.46){\Blue{\line(1,-1){0.5}}}
\put(4.08,2.51){\Blue{\line(1,-1){0.5}}}
\put(4.13,2.56){\Blue{\line(1,-1){0.5}}}
\put(4.18,2.61){\Blue{\line(1,-1){0.5}}}
\put(4.23,2.66){\Blue{\line(1,-1){0.5}}}
\put(4.28,2.71){\Blue{\line(1,-1){0.5}}}
\put(4.33,2.76){\Blue{\line(1,-1){0.5}}}
\put(4.38,2.81){\Blue{\line(1,-1){0.5}}}
\put(4.43,2.86){\Blue{\line(1,-1){0.5}}}
\put(4.48,2.91){\Blue{\line(1,-1){0.5}}}
\put(4.58,2.91){\Blue{\line(1,-1){0.45}}}
\put(4.68,2.91){\Blue{\line(1,-1){0.40}}}
\put(4.78,2.91){\Blue{\line(1,-1){0.35}}}
\put(4.88,2.91){\Blue{\line(1,-1){0.30}}}
\put(4.98,2.91){\Blue{\line(1,-1){0.25}}}
\put(5.08,2.91){\Blue{\line(1,-1){0.20}}}
\put(5.18,2.91){\Blue{\line(1,-1){0.15}}}
\put(5.28,2.91){\Blue{\line(1,-1){0.10}}}
\put(3.24,1.96){$p$}
\put(4.65,1.96){$q$}
\end{picture}
\caption{Example 6, cylinder universe with degenerate regions.}
\label{figure 7}
\end{center}
\end{figure}
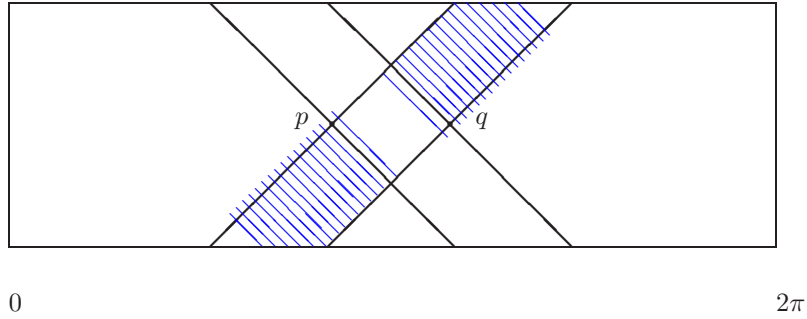
\\* \\*
The above example is very unfortunate, in the sense that it shows that using
only relations between points derived from the chronological partial
order on zero-dimensional objects is not sufficient for obtaining a
satisfactory causal relation.  However, it also suggests that the following
definition might be more successful.
\begin{deffie}
Define a partial order $\mathcal{P}$ as $p\, \mathcal{P}\, q$ iff 
there exists a
continuous curve $\gamma : \left[0,1\right] \rightarrow \mathcal{M}$ from $p$
to $q$ such that $\gamma(t)\, \mathcal{R}\, \gamma(s)$ for all $0 \leq t\leq s
\leq 1$. Finally, define $\leq_{d}$ on $\mathcal{M} \setminus
\overline{\mathcal{TCON}}$ as the smallest topologically closed transitive
relationship extending $\mathcal{P}$ and $I^{+}$.  It is easy to see that
$\leq_{d}$ is compatible with $K^{+}$, i.e., $p \leq_{d} q$ and $q \in
K^{-}(r)$ imply that $p \in K^{-}(r)$ and vice versa.
\end{deffie}

The following example in three dimensions shows that this definition also has
its limitations.  However, in two dimensions it does work, as is proven in
Theorem $7$.
\begin{example}
\end{example}
Consider the three-dimensional cylinder universe $({\rm S}^2 \times
[-1,1], -\dd t^2 + \dd\theta^2 + \sin^2\!\theta\,\dd\phi^2)$, and the
spacelike geodesic in it defined by $\gamma: [0,\frac{1}{4} ]
\rightarrow {\rm S}^2 \times [-1,1]: s \rightarrow \gamma(s)
= (\theta_0 +s , \phi_0, 0)$.  Take the limit $({\rm S}^2 \times
[-1,1],d)$ over a suitable sequence of conformally equivalentå
metrics, with conformal factors which converge to zero on thin specific (see
figure 8) open neighborhoods of $J^+(\gamma(s)) \setminus
J^{+}(\gamma(t)) $ and $J^{-}(\gamma(t)) \setminus J^{-}(\gamma(s))$ which are
subsets of $J^{+}(\gamma(t))^{\rm c}$ and $J^{-}(\gamma(s))^{\rm c}$,
respectively, for all $t<s$.\footnote{The relation $J^{+}$ denotes here the
usual causal relation defined by the line element $-\dd t^2 + \dd\theta^2 +
\sin^2\!\theta\,\dd\phi^2$.}
\begin{figure}[h]
\begin{center}
  \setlength{\unitlength}{1.7cm}
\begin{picture}(8,3.5)

\put(1,1){\line(1,0){6}}
\put(1,1){\line(0,1){1.91}}
\put(1,2.91){\line(1,0){6}}
\put(7,1){\line(0,1){1.91}}
\put(1,0.5){$0$}
\put(7,0.5){$2 \pi$}
\put(3.53,1.96){\circle*{0.05}}
\put(4.45,1.96){\circle*{0.05}}
\thicklines
\put(3.53,1.96){\line(1,0){0.94}}
\put(3.53,1.96){\circle{1}}
\put(4.45,1.96){\circle{1}}
\put(3.53,2.37){\line(1,0){0.94}}
\put(3.53,1.54){\line(1,0){0.94}}
\thinlines
\put(3.53,2.42){\Blue{\line(1,0){0.99}}}
\put(3.64,2.38){\Blue{\line(1,0){0.99}}}
\put(3.75,2.32){\Blue{\line(1,0){0.99}}}
\put(3.80,2.27){\Blue{\line(1,0){0.99}}}
\put(3.85,2.22){\Blue{\line(1,0){0.99}}}
\put(3.89,2.17){\Blue{\line(1,0){0.99}}}
\put(3.91,2.12){\Blue{\line(1,0){0.99}}}
\put(3.92,2.07){\Blue{\line(1,0){0.99}}}
\put(3.94,2.02){\Blue{\line(1,0){0.99}}}
\put(3.95,1.97){\Blue{\line(1,0){0.99}}}
\put(3.94,1.92){\Blue{\line(1,0){0.99}}}
\put(3.92,1.87){\Blue{\line(1,0){0.99}}}
\put(3.91,1.82){\Blue{\line(1,0){0.99}}}
\put(3.89,1.77){\Blue{\line(1,0){0.99}}}
\put(3.87,1.72){\Blue{\line(1,0){0.99}}}
\put(3.83,1.67){\Blue{\line(1,0){0.99}}}
\put(3.79,1.62){\Blue{\line(1,0){0.99}}}
\put(3.73,1.57){\Blue{\line(1,0){0.99}}}
\put(3.63,1.52){\Blue{\line(1,0){0.99}}}
\put(3.52,1.47){\Blue{\line(1,0){0.99}}}
\put(3.24,1.96){$p$}
\put(4.65,1.96){$q$}
\end{picture}
\caption{Example 7, lightcones at fixed time $t > 0$ drawn on a Euclidean
coordinate patch.  The shading indicates the degenerate regions.}
\label{figure 8}
\end{center}
\end{figure}
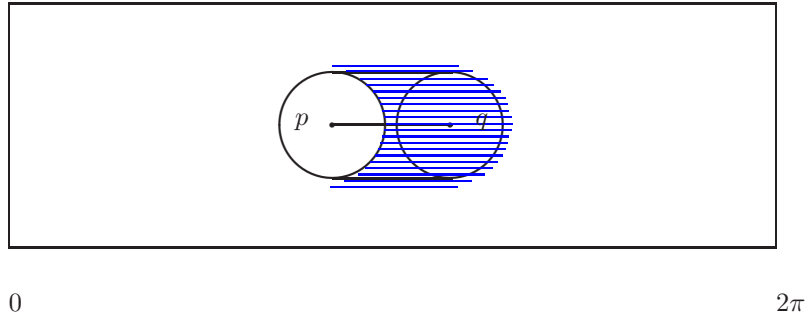
It is not difficult to see that for any point $q$ belonging to the degenerate
region, either $I_{d}^{+}(q) \cap I_{d}^{+}(\gamma(0)) \neq \emptyset$ or
$I_{d}^{-}(q) \cap I_{d}^{-}(\gamma(1/4)) \neq \emptyset$.  Using this, it is
easy to see that for any two points $p$ and $q$ belonging to the degenerate
region, we have that $I_{d}^{-}(p) \neq I_{d}^{-}(q)$ or $I_{d}^{+}(p) \neq
I_{d}^{+}(q)$. \hfill$\square$  \\*

\begin{theo} Let $(\mathcal{M},g)$ be a \emph{two-dimensional} globally
hyperbolic interpolating spacetime wich is isometrically embeddable in the
interior of an interpolating spacetime \emph{without cut points} and suppose
$\Omega_{i}$ is a sequence of positive $C^{\infty}$ functions on $\mathcal{M}$
such that $\big| d_{\Omega_{i}^{2}g}(p,q) - d_{\Omega_{j}^{2}g}(p,q) \big|
< \frac{1}{i}$ for all $j > i > 0$ and $p,q \in \mathcal{M}$.  Denote by
$(\mathcal{N},d)$ the GGH limit space \emph{and} suppose that
$\mathcal{M}=\mathcal{N}$, i.e., no points get identified.  Then, one has that
$p \leq_{g} q$ iff $p \leq_{d} q$ for all $p,q	\in \mathcal{M}$.
\end{theo}
\textbf{Note}: It is easy to see that if $(\man,g)$ were allowed to have cut
points, then the theorem would not be valid anymore.  Readers should convince
themselves of this by making a drawing on the two dimensional cylinder
universe.
\\*
\\*
\textsl{Proof:}
\\*
First notice that the strong topology defined by $D$ coincides with 
the manifold
topology.
\\*
$\Rightarrow)$ We show that \emph{any} $g$-causal curve $\gamma$
is an $\mathcal{R}$-causal curve. Clearly, $I^{+}_{d}(\gamma(s))
\subseteq I^{+}_{d}(\gamma(t))$ and $I^{-}_{d}(\gamma(t))
\subseteq I^{-}_{d}(\gamma(s))$ for all $t < s$, which proves the
basis of induction.  Let $\alpha = \beta + 1$ and suppose that $t
< s$ implies that $\gamma(s) \prec^\beta y \Rightarrow
\gamma(t) \prec^\beta y$.  Obviously, if $\gamma(s)
\prec^\beta y \prec^{\beta} z$ then $\gamma(t) \prec^\beta y
\prec^\beta z$.  So, suppose that there exist sequences
$\{q_{n}\}_{n\in{\bf N}}$ and $\{y_{n}\}_{n\in{\bf N}}$,
converging to $\gamma(s)$ and $y$ respectively, such that $q_{n}
\prec^{\beta} y_{n}$ for any $n$; then, since $(\mathcal{M},g)$
has no cut points, there exists a sequence $\{p_{n}\}_{n\in
{\bf N}}$ converging to $\gamma(t)$ with $p_{n} \prec_{g}
q_{n}$.  The induction hypothesis then implies that $p_{n}
\prec^{\beta} y_{n}$ for all $n$, which proves the claim.
\\*
$\Leftarrow)$ We have to show that $g$-spacelike events cannot be 
connected by an
$\mathcal{R}$-causal curve.  Suppose that $p$ and $q$ are such events, and that
$\gamma: \left[0,1\right] \rightarrow \mathcal{M}$ is an $\mathcal{R}$-causal
curve between them.  Without loss of generality, we may assume that $\gamma$ is
spacelike related to $p$ and $q$, in the sense that $\gamma(t)$, $p$ 
and $q$ are
$g$-spacelike events for all $t \in (0,1)$.\footnote{Note that $\gamma$
cannot intersect $E^{-}(p)$ nor $E^{+}(q)$, because this would violate the
assumtion that the limit space equals $\mathcal{M}$.  By continuity, 
there exists
a maximal $t$ and a minimal $s>t$ such that $\gamma(t) \in E^{+}(p)$, but
$\gamma(u) \notin E^{+}(p)$ for all $u > t$ and $s$ is the minimal 
number larger
than $t$ such that $\gamma(s) \in E^{-}(q)$.}  Moreover, we may assume that
$\gamma$ is a subset of a convex open neighborhood $\mathcal{U}$ on 
which $g$ is
conformally flat.  By using the nonexistence of cut points, one can deduce that
the set $\mathcal{S} \subset \mathcal{M}$ bounded by the two right $g$ null
geodesics containing $\gamma$ is entirely degenerate, as shown in figure
$9$.\footnote{Choose $\gamma(t), t \in \left(0,1\right)$.  Then there exists an
open neighborhood $\mathcal{O}$ of $\gamma(t)$ such that for all $r \in
\mathcal{O}$ which are $g$ spacelike to the left or in the $g$ 
chronological past
of $\gamma(t)$, we have that the left, future oriented, null geodesic 
starting at
$r$ does \emph{not} intersect the future oriented, right null 
geodesic starting at
$\gamma(t)$ otherwise $\gamma(t)$ would have a cut point in any extension of
$(\mathcal{M},g)$.  Hence for all $s < t$ such that $\gamma(s) \in 
\mathcal{O}$ is
such point $r$, we have that any point in $J^{+}(\gamma(t))$ to the 
right of the
right null geodesic emanating from $\gamma(s)$ belongs to the 
degenerate area.  A
similar argument is valid for the past with left and right switched. 
Using this
for all $t$ leads to picture $9$.}  Any two points in $\mathcal{S} \cap
\mathcal{U}$ belonging to any left $g$ null geodesic have the same 
chronological
relations.
\begin{figure}[h]
\begin{center}
  \setlength{\unitlength}{1.7cm}
\begin{picture}(7,3)
\put(1,1){\line(1,0){6}}
\put(1,1){\line(0,1){1.91}}
\put(1,2.91){\line(1,0){6}}
\put(7,1){\line(0,1){1.91}}
\put(3.53,1.96){\circle*{0.05}}
\put(4.45,1.96){\circle*{0.05}}
\thicklines
\put(3.53,1.96){\line(1,1){0.95}}
\put(3.53,1.96){\line(1,-1){0.95}}
\put(3.53,1.96){\line(-1,1){0.95}}
\put(3.53,1.96){\line(-1,-1){0.95}}
\put(4.45,1.96){\line(1,1){0.95}}
\put(4.45,1.96){\line(1,-1){0.95}}
\put(4.45,1.96){\line(-1,1){0.95}}
\put(4.45,1.96){\line(-1,-1){0.95}}
\put(3.53,1.96){\line(1,0){0.95}}
\put(3.8,1.3){\line(-1,1){0.3}}
\thinlines
\put(3.58,2.01){\Blue{\line(1,-1){0.45}}}
\put(3.48,1.91){\Blue{\line(1,-1){0.45}}}
\put(3.43,1.86){\Blue{\line(1,-1){0.45}}}
\put(3.38,1.81){\Blue{\line(1,-1){0.45}}}
\put(3.33,1.76){\Blue{\line(1,-1){0.45}}}
\put(3.28,1.71){\Blue{\line(1,-1){0.45}}}
\put(3.23,1.66){\Blue{\line(1,-1){0.45}}}
\put(3.18,1.61){\Blue{\line(1,-1){0.45}}}
\put(3.13,1.56){\Blue{\line(1,-1){0.45}}}
\put(3.08,1.51){\Blue{\line(1,-1){0.45}}}
\put(3.03,1.46){\Blue{\line(1,-1){0.45}}}
\put(2.98,1.41){\Blue{\line(1,-1){0.4}}}
\put(2.93,1.36){\Blue{\line(1,-1){0.35}}}
\put(2.88,1.31){\Blue{\line(1,-1){0.30}}}
\put(2.83,1.26){\Blue{\line(1,-1){0.25}}}
\put(2.78,1.21){\Blue{\line(1,-1){0.20}}}
\put(2.73,1.16){\Blue{\line(1,-1){0.15}}}
\put(2.68,1.11){\Blue{\line(1,-1){0.10}}}
\put(2.63,1.06){\Blue{\line(1,-1){0.10}}}
\put(2.58,1.01){\Blue{\line(1,-1){0.10}}}

\put(3.48,1.91){\Blue{\line(1,-1){0.45}}}
\put(3.53,1.96){\Blue{\line(1,-1){0.45}}}
\put(3.58,2.01){\Blue{\line(1,-1){0.45}}}
\put(3.63,2.06){\Blue{\line(1,-1){0.45}}}
\put(3.68,2.11){\Blue{\line(1,-1){0.45}}}
\put(3.73,2.16){\Blue{\line(1,-1){0.45}}}
\put(3.78,2.21){\Blue{\line(1,-1){0.45}}}
\put(3.83,2.26){\Blue{\line(1,-1){0.45}}}
\put(3.88,2.31){\Blue{\line(1,-1){0.45}}}
\put(3.93,2.36){\Blue{\line(1,-1){0.45}}}
\put(4.03,2.46){\Blue{\line(1,-1){0.45}}}
\put(4.08,2.51){\Blue{\line(1,-1){0.45}}}
\put(4.13,2.56){\Blue{\line(1,-1){0.45}}}
\put(4.18,2.61){\Blue{\line(1,-1){0.45}}}
\put(4.23,2.66){\Blue{\line(1,-1){0.45}}}
\put(4.28,2.71){\Blue{\line(1,-1){0.45}}}
\put(4.33,2.76){\Blue{\line(1,-1){0.45}}}
\put(4.38,2.81){\Blue{\line(1,-1){0.45}}}
\put(4.43,2.86){\Blue{\line(1,-1){0.45}}}
\put(4.48,2.91){\Blue{\line(1,-1){0.45}}}
\put(4.58,2.91){\Blue{\line(1,-1){0.40}}}
\put(4.68,2.91){\Blue{\line(1,-1){0.35}}}
\put(4.78,2.91){\Blue{\line(1,-1){0.30}}}
\put(4.88,2.91){\Blue{\line(1,-1){0.25}}}
\put(4.98,2.91){\Blue{\line(1,-1){0.20}}}
\put(5.08,2.91){\Blue{\line(1,-1){0.15}}}
\put(5.18,2.91){\Blue{\line(1,-1){0.10}}}
\put(5.28,2.91){\Blue{\line(1,-1){0.10}}}
\put(3.24,1.96){$p$}
\put(4.65,1.96){$q$}
\end{picture}
\caption{Proof of theorem 7, conflict with the $T_{0}$ property.}
\label{figure 9}
\end{center}
\end{figure}
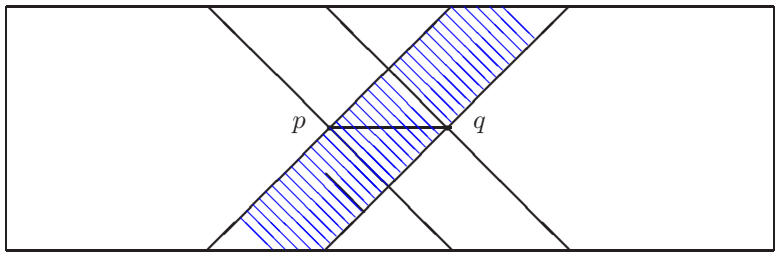
\hfill$\square$
\\*
\\*
The results of Example 8 and Theorem 7 are quite discouraging, since any good
definition seems to depend upon some notion of \emph{dimension} of the Lorentz
space.  One could try to make the definition more restrictive so that it would
be possible to reproduce a result analogous to Theorem 7 in all dimensions.
It seems to us that ``local"\footnote{Local in the sense that one uses
properties of local congruences of curves between neighborhoods of points.
One such idea would be to construct the following kind of definition: define
the relation $p\, \mathcal{Q}\, q$ iff there exist neighborhoods $\mathcal{U}$,
$\mathcal{V}$ of $p$ and $q$, respectively, and a mapping $\psi: \mathcal{U}
\times \left[0,1 \right] \rightarrow \mathcal{M}$ such that
\begin{itemize}
\item $\psi_{t} : \mathcal{U} \rightarrow \mathcal{M}: r \rightarrow
\psi(r,t)$ is a homeomorphism for any $t$ and $\psi_{1} (\mathcal{U}) =
\mathcal{V}$.
\item $\psi_{r}: \left[0,1\right] \rightarrow \mathcal{M} : t \rightarrow
\psi(r,t)$ defines a $\mathcal{R}$-causal curve for any $r \in \mathcal{U}$.
\end{itemize}
$\leq_{d}$ is then defined as the smallest topologically closed transitive
relation encompassing $\mathcal{Q}$ and $I^{+}$.  Again it is not difficult
to construct a counterexample similar to example 7.} ideas won't work; we are
working on other ideas concerning a promising definition, but we do not have
any proof yet.
\\*
\\*
The rest of this section is devoted to proving that the limit space
$(\mathcal{M},d)$ of a $\mathcal{C}^+_\alpha$ and $\mathcal{C}^-_\alpha$
sequence  $\{(\mathcal{M}_{i}, d_{i})\}_{i\in\NN}$ of path metric\footnote{For
a precise definition of a path metric Lorentz space, see Definition 7.} Lorentz
spaces is a path metric Lorentz space.  Strictly speaking, we should 
still define
the $\mathcal{C}^{\pm}_{\alpha}$ properties for general Lorentz spaces
$(\mathcal{M},d)$.  Looking at the definition in the Intermezzo of section 3 of
Ref \cite{Noldus2}, the reader can see that this boils down to defining the
future and past boundaries of $(\mathcal{M},d)$.  Obviously, $\partial_{\rm P}
\mathcal{M}$ is the set of points $p$ such that $I^{-}(p) = \emptyset$, and
$\partial_{\rm F}\mathcal{M}$ is defined dually.
\\*
\\*
\textbf{Property}: The $\mathcal{C}^+_\alpha$ property implies that the
interior of $\partial_{\rm P} \mathcal{M}$ is empty and, likewise, the
$\mathcal{C}^-_\alpha$ property implies that the interior of $\partial_{\rm F}
\mathcal{M}$ is empty.
\\*
\\*
\textsl{Proof}:
\\*
We will only prove the first claim. Notice that $\partial_{\rm P} \mathcal{M}
\cap \partial_{\rm F} \mathcal{M}$ contains at most one point.  Let $p \in
\partial_{\rm P} \mathcal{M} \setminus \partial_{\rm F} \mathcal{M}$ 
and $\epsilon
>  0$ be such that $B_{D}(p, \epsilon) \subset \partial_{\rm P} 
>\mathcal{M} \setminus
\partial_{\rm F} \mathcal{M}$.  Then $d(p,r) = 0$ for all $r \in B_{D}(p,
\epsilon)$, which is impossible by the $\mathcal{C}^+_\alpha$ property.
\hfill$\square$
\\*
\\*
As a consequence, we have that for a Lorentz space $(\mathcal{M},d)$ 
satisfying the
$\mathcal{C}^+_\alpha$ and $\mathcal{C}^{-}_{\alpha}$ property,
$\overline{\mathcal{TCON}} \cup \left( \partial_{\rm P} \mathcal{M} \cap
\partial_{\rm F} \mathcal{M} \right) = \mathcal{M}$.\footnote{For 
example, let $p
\in \partial_{\rm P} \mathcal{M} \setminus \partial_{\rm F} 
\mathcal{M}$; then for
$\epsilon > 0$ sufficiently small, we have that $B_{D}(p, \epsilon) \cap
\partial_{\rm F} \mathcal{M} = \emptyset$. By the 
$\mathcal{C}^+_\alpha$ property,
there exists an $r \in \overline{B_{D}(p,\frac{\epsilon}{2})}$ such 
that $d(p,r) =
\alpha(\frac{\epsilon}{2})$. Hence,
$$
    D(r,\partial_{\rm P} \mathcal{M}), D(r, \partial_{\rm F} \mathcal{M})
    \geq \alpha\, (\epsilon/2)\;,
  $$
which implies, by the $\mathcal{C}^+_\alpha$ and $\mathcal{C}^{-}_{\alpha}$
properties, that $r \in \mathcal{TCON}$.}  Note that the second term on the
left-hand side of this equality only needs to be accounted for iff 
$\partial_{P}
\mathcal{M} \cap \partial_{\rm F} \mathcal{M}$ is an \emph{isolated} 
point.  Hence,
the causal relation on such space is the $K^{+}$ relation.
\\*
\\*
First, we give an example of spacetime with the $\mathcal{C}^{\pm}_{x^{2}/2}$
property.
\begin{example} \label{ex18}
\end{example}
Consider again the cylinder universe $\mathcal{CYL} = ({\rm S}^1
\times  [0,1], -\dd t^{2} + \dd\theta^2)$.  We argue that $\mathcal{CYL}$
belongs to the  category defined by $\alpha : \RR^{+} \rightarrow \RR^{+} : x
\rightarrow \half\,x^{2}$.  Since SO(2) is the isometry group of $d$, it is
sufficient to prove the assertion for  points with a fixed spatial coordinate,
say $\theta = \pi$.  Let $1 \geq \tilde{t} > t \geq 0$; then it is 
easy to prove
that
$$
        D((\pi , t) ,(\pi ,\tilde{t})) = \sqrt{\tilde{t} - t} \,
        \max \Big\{\sqrt{\tilde{t}+t} , \sqrt{2- (t+\tilde{t}) } \Big\},
$$
where $D$ is the strong metric on $\mathcal{CYL}$.  Hence
$$
        d((\pi,t) , (\pi, \tilde{t})) \leq D((\pi , t) ,(\pi ,\tilde{t}))^{2}
        \leq 2\, d((\pi,t) , (\pi, \tilde{t}))\;,
$$
which proves the assertion. \hfill$\square$
\\*
\\*
Before we proceed, we should define causal curves $\gamma$ and lengths
thereof.\footnote{The reader can find similar definitions in Ref 
\cite{Busemann}.}
\begin{deffie}
Let $(\man,d)$ be a Lorentz space, assume that $a<b$ and let $\gamma : [a, b]
\rightarrow \mathcal{M}$ be a continuous mapping (with respect to the strong
topology) such that for all $a \leq t < s \leq b$, $ \gamma (t) \leq 
\gamma (s)$
($\gamma (t) \ll \gamma (s)$); then $\gamma$ is a basic causal 
(timelike) curve.
Now let $a < b$, $c < d$ and $\gamma_{1}: [a, b] \rightarrow \mathcal{M}$,
$\gamma_{2}: [c ,d] \rightarrow \mathcal{M}$ be causal curves such 
that $\gamma_2(c)
= \gamma_1(b)$; we define the concatenation $\gamma_2 \circ \gamma_1$ 
of $\gamma_2$
with $\gamma_{1}$ as the basic causal curve $\gamma_2 \circ \gamma_1: 
[a, b + d -
c] \rightarrow \mathcal{M}$ such that
$$
        \gamma_2 \circ \gamma_1(t)
        = \cases{ \gamma_1(t) & {\rm if} $a \leq t \leq b$ \cr
        \gamma_2 (t + c - b) & {\rm if} $b \leq t \leq b + d - c$.}
$$
A (countably infinite) concatenation of basic causal curves is a causal
curve.
\end{deffie}

The length $L(\gamma)$ of a basic causal curve $\gamma : [a,b] \rightarrow
\mathcal{M}$ is defined as
$$
        L(\gamma) = \inf_{ \Delta } \sum_{i = 0}^{ \left| \Delta \right| - 1}
        d(\gamma(t_{i}) , \gamma(t_{i+1}))\;,
$$
where $\Delta = \left\{t_{i} \mid a = t_{0} < t_{1} < \ldots < t_{n-1} < t_n
= b \right\}$ is a partition of $[a,b]$.  Obviously,
$$
        L(\gamma_{2} \circ \gamma_{1}) = L(\gamma_{1}) + L(\gamma_{2})\;.
$$

Now, we are able to give the definition of a path metric Lorentz space.
\begin{deffie}
$(\man,d)$ is a path metric lorentz space iff for any $p \leq q$ there exists
a causal curve $\gamma$ from $p$ to $q$ such that $L(\gamma) = d(p,q)$.
\end{deffie}
In case the causal relation coincides with the $K^+$ relation, we can 
prove that
$(\mathcal{M},d)$ is a path metric space iff for any $p \ll q$, there exists a
distance-realising ($K^+$) causal curve from $p$ to $q$.  We need to 
introduce the
Vietoris topology on the set $2^{(\mathcal{M},D)}$ of all closed, 
non-empty subsets
of $(\mathcal{M},D)$ for which a \emph{sub-basis} is given by the sets
$\mathcal{B}(\mathcal{M},\mathcal{O})$ and $\mathcal{B}(\mathcal{O}, 
\mathcal{M})$.
The former are sets with as members closed sets which meet the open set
$\mathcal{O}$, the latter consists of the closed subsets of 
$\mathcal{O}$.  It is
known that $2^{(\mathcal{M},D)}$ equipped with the Vietoris topology is compact
\cite{SW}; in that reference, it is also proven using topological 
arguments only
that the Vietoris limit of a sequence of $K^+$-causal curves is a $K^+$-causal
curve.
\begin{theo}
\label{theo27} Let $(\mathcal{M},d)$ be a Lorentz space; then $(\mathcal{M},d)$
is a path metric space with respect to the $K^{+}$ relation iff for 
any $p \ll q$,
there exists a distance-realising $K^+$-causal curve from $p$ to $q$.
\end{theo}
\textsl{Proof}: \\* We only have to prove that the latter implies
the former, the other way around being obvious.  We shall once
more proceed by transfinite induction, using as induction hypothesis
$H_{\alpha}$ the statement that $p \prec^\alpha q$ implies that
there exists a distance-realising $K^+$-causal curve from $p$ to
$q$.  The basis of induction is nothing else but our assumption.
Hence, let $\alpha = \beta + 1$ and assume $H_{\beta}$ is valid.
If $p \prec^{\beta} q \prec^{\beta} r$, then there exists a $K^+$-causal
curve from $p$ to $r$, by the induction hypothesis and concatenation,
which is obviously distance-maximising.  So assume that there exist sequences
$\{p_{n}\}_{n \in {\bf N}}$ and $\{q_{n}\}_{n \in {\bf N}}$ 
converging to $p$ and
$q$, respectively, and that $p_{n} \prec^{\beta} q_{n}$ for all $n \in
{\bf N}$.  Then, $H_{\beta}$ implies that there exist $K^+$-causal curves
$\gamma_{n}$ from $p_{n}$ to $q_{n}$.  We may assume that, by passing to a
subsequence if necessary, $\{\gamma_{n}\}_{n \in {\bf N}}$ converges in the
Vietoris topology to a (distance-maximising) $K^+$-causal curve connecting $p$
with $q$.
\hfill$\square$
\\*
\\*
Before we turn to the study of properties of the space of causal curves
between two points, it is necessary to look at some properties of causal
curves  with respect to the strong metric $D$.
\begin{example}
\end{example}
We show that the $D$-length of a compact, basic causal curve is in general
infinity. Obviously, the way to define the $D$-length, $DL(\gamma)$,
of a basic causal curve $\gamma: [a, b] \rightarrow \mathcal{M}$ is
$$
        DL( \gamma ) = \sup_{\Delta} \sum_{i = 0}^{\left| \Delta \right| - 1 }
        D(\gamma (t_{i}) , \gamma(t_{i+1}))\;,
$$
where, as before, $\Delta = \left\{t_{i} \mid a = t_{0} < t_{1} <
\ldots < t_{n-1} < t_{n} = b \right\}$ is a partition of $[a,b]$.
Returning to Example $1$, to prove that the length of the interval $\left\{
(\pi , t) \mid 0 \leq t \leq 1 \right\}$ equals $\infty$ we choose $\Delta_{n}
= \left\{ 1 - \frac{1}{k} \mid k  = 1 \ldots n \right\} \cup \{1\}$; for $k>2$
$$
     D((\pi,t_k),(\pi,t_{k+1})) = D^-((\pi,t_k),(\pi,t_{k+1}))
     = {\sqrt{2k^2-1}\over k\,(k+1)} > {1\over k}
$$
and the sum diverges, which proves the claim. \hfill$\square$
\\*
\\*
Since the above example shows that the $D$-length of a causal curve  is a
meaningless concept, we have to come up with some other way to divide a
causal curve into smaller pieces.  As a starter, we mention the following
result.
\begin{theo}
Let $\gamma : \left[a , b \right] \rightarrow \mathcal{M}$ be any
basic causal curve in an interpolating spacetime $(\man,g)$. Then there
exists no $t \in (a,b)$ such that $D(\gamma(a), \gamma(t)) = D(\gamma(t)
,\gamma(b)) = \half\,D(\gamma(a), \gamma(b))$, i.e., $\gamma$ has no
$D$-midpoint.
\end{theo}
\textsl{Proof:}
\\*
We show that for all $t \in \left( a ,b \right)$: $D(\gamma(a),
\gamma(b)) < D(\gamma(a) , \gamma(t)) + D(\gamma(t), \gamma(b))$.
Assume that the point $r$, which realizes $D(\gamma(a), \gamma(b))$,
belongs to $I^{+} (\gamma(a)) \setminus I^{+} (\gamma(b))$; the case
where $r$ belongs to $I^{-}(\gamma(b)) \setminus I^{-} (\gamma(a))$
is identical and is left as an exercise to the reader.  Hence,
$$
    D(\gamma(a), \gamma(b)) =  d(\gamma(a),r) = \left( d(\gamma(a),r)
    - d(\gamma(t),r) \right) + d(\gamma(t),r)\;.
$$
Both terms on the rhs are non-negative, and bounded by $D(\gamma(a) ,
\gamma(t))$ and $D(\gamma(t) , \gamma(b))$ respectively.  The first
term can only realize $D(\gamma(a), \gamma(t))$ if $r \in
I^{+}(\gamma(a)) \setminus I^{+} (\gamma(t))$.  But in that case
$d(\gamma(t),r) = 0$, which concludes the proof.  \hfill$\square$
\\*
\\*
Hence, we define the following concept of division of a causal curve.
\begin{deffie}
Let $\gamma : \left[ a ,b \right] \rightarrow \mathcal{M}$ be any basic
causal curve, and denote $\delta = D (\gamma(a) , \gamma(b))$.  The division
of $\gamma$ is the set of points $\gamma^{1/2} = \left\{p_{i} \mid i=0
\ldots k \right\}$ such that $\gamma (a) = p_{0} \leq p_{1} \prec \ldots \prec
p_{k-1} \prec p_{k} = \gamma (b)$, $D (p_{i} , p_{i+1}) =
\textstyle{\delta\over2},\; \forall i: 0 \ldots k-2$, $\textstyle{\delta\over4}
\leq D(p_{k-1}, p_{k}) < {\textstyle{3\delta\over4}}$ and there exists no point
$q$ such that $p_{k-1} \prec q \prec \gamma(b)$ with $D(p_{k-1},q)
=\textstyle{\delta\over2}$ and $\textstyle{\delta\over4} \leq D(q, \gamma(b)) <
{\textstyle{3\delta\over4}}$. Such finite number $2 \leq k = 
|\gamma^{1/2}| - 1$
exists, since $\gamma$ is continuous with respect to the strong topology.

\end{deffie}
This new concept facilitates the proof of the final theorem.
\begin{theo}
The limit space $(\mathcal{M},d)$ of a GGH $\mathcal{C}^{+}_{\alpha}$ and
$\mathcal{C}^{-}_{\alpha}$ Cauchy sequence $\{(\man_i, d_i)\}_{i\in\NN}$
of path metric Lorentz spaces, is a path metric Lorentz space.
\end{theo}
\noindent\textsl{Proof:}
According to Theorem \ref{theo27} and the arguments preceding it, we only have
to prove that $p \ll q$, $p,q \in \mathcal{M}$, implies that there 
exists a $K^+$
causal curve connecting $p$ with $q$.
Let $\psi_{i}:\mathcal{M}_{i} \rightarrow \mathcal{M}$ and $\zeta_{i}:
\mathcal{M} \rightarrow \mathcal{M}_{i}$ be mappings under which
$(\mathcal{M}_{i},d_{i})$ and $(\mathcal{M},d)$ are
$(\epsilon_{i},\epsilon_{i})$-close, where
$\epsilon_{i} \stackrel{i \rightarrow \infty}{\rightarrow} 0$.
Choose $p,q \in \mathcal{M}$ such that $d(p,q) > 0$.
Choose $\epsilon < \frac{1}{8}\, d(p,q)$ and choose $i$
sufficiently large, such that $\epsilon_{i} < \alpha(\epsilon)$.  Hence,
$\left| d_{i}(\zeta_{i}(p), \zeta_{i}(q)) - d(p,q) \right| < \epsilon_{i}$
and $\left| D_{i}(\zeta_{i}(p), \zeta_{i}(q)) - D(p,q)  \right| < 4\,
\epsilon_{i}$.  Let $\gamma_{i}$, be a geodesic from $\zeta_{i}(p)$ to
$\zeta_{i}(q)$ and consider $\gamma_{i}^{1/2} = \left\{p^{i}_{s} \mid s=0
\ldots k_{i} \right\}$.  Assume that $s^{i}$ is the largest number such that
$d_{i}(p^{i}_{s^{i}+1}, \zeta_{i}(q)) > 
\half\,d_{i}(\zeta_{i}(p),\zeta_{i}(q))$.
Then, for all $s \leq s^{i}$, choose $r^{i}_{s+1}$ such that
$\zeta_{i}(q) \gg_i r^{i}_{s+1} \gg_i p^{i}_{s+1}$,
$D_{i}(p^{i}_{s+1},r^{i}_{s+1}) \le \epsilon$ and
$d_{i}(p^{i}_{s+1},r^{i}_{s+1}) = \alpha (\epsilon )$.  This is possible,
since the $\mathcal{C}^{+}_{\alpha}$ property is valid and since
$\half\, d_{i}(\zeta_{i}(p),\zeta_{i}(q)) >
\half\, [d(p,q)-\alpha(\epsilon)] > \frac{7}{16}\,d(p,q)$.  If
$d_{i}(p^{i}_{s^{i}+1},p^{i}_{s^{i}+2}) < \alpha(\epsilon)$, then construct in
a similar way $r^{i}_{s^{i}+2}$; this is possible since $\frac{5}{16}\,d(p,q) >
\epsilon$.  For all $s > s^{i}+1$, define $t^{i}_{s} \ll_i p^{i}_{s}$ such
that $D_{i}(t^{i}_{s},p^{i}_{s+1}) \le \epsilon$ and
$d_{i}(t^{i}_{s},p^{i}_{s+1}) = \alpha (\epsilon)$.  Obviously,
$d_{i}(\zeta_{i}(p),t^{i}_{s}),d_{i}(r^{i}_{s},\zeta_{i}(q)) >
\frac{3}{16}\,d(p,q)$ and $d_{i}(p^{i}_{s},r^{i}_{s+1}), d_{i}(t^{i}_{s},
p^{i}_{s+1}) > \alpha \left( \epsilon \right)$.  Hence, one can uniquely
define sequences of the types
$$
    \left\{\zeta_{i}(p),r^{i}_{1},p^{i}_{1},
    r^{i}_{2}, p^{i}_{2}, \ldots , p^{i}_{s^{i}}, r^{i}_{s^{i}+1},
    p^{i}_{s^{i}+1}, r^{i}_{s^{i}+2}, t^{i}_{s^{i}+2}, p^{i}_{s^{i}+3},
    t^{i}_{s^{i}+3}, \ldots , t^{i}_{k_{i}-1}, \zeta_{i}(q) \right\}
$$
and
$$
    \left\{\zeta_{i}(p),r^{i}_{1},p^{i}_{1}, r^{i}_{2}, p^{i}_{2}, \ldots ,
    p^{i}_{s^{i}}, r^{i}_{s^{i}+1}, p^{i}_{s^{i}+1}, p^{i}_{s^{i}+2},
    t^{i}_{s^{i}+2}, p^{i}_{s^{i}+3}, t^{i}_{s^{i}+3}, \ldots , t^{i}_{k_{i}-1},
    \zeta_{i}(q) \right\}
$$
depending on whether
$d_{i}(p^{i}_{s^{i}+1},p^{i}_{s^{i}+2}) < \alpha(\epsilon)$ or
$d_{i}(p^{i}_{s^{i}+1},p^{i}_{s^{i}+2}) \geq \alpha(\epsilon)$, respectively.
In general, we have constructed a sequence of the form
$\{z^{i}_{s}\}_{s=0}^{2k_{i}-1}$ with the following useful properties:
\begin{itemize}
\item $z^{i}_{0} = \zeta_{i}(p)$ and $z^{i}_{2k_{i}-1} = \zeta_{i}(q)$;
\item $\half\, D_{i}(\zeta_{i}(p),\zeta_{i}(q)) \leq D_{i}(z^{i}_{2s},
z^{i}_{2s+1}) \leq \half\, D_{i}(\zeta_{i}(p),\zeta_{i}(q)) + \epsilon$
for $s \leq k^{i}-2$,
\\*
$\frac{1}{4}\, D_{i}(\zeta_{i}(p),\zeta_{i}(q)) \leq D_{i}(z^{i}_{2k^{i}-2},
z^{i}_{2k^{i}-1}) < \frac{3}{4}\, D_{i}(\zeta_{i}(p),\zeta_{i}(q)) + \epsilon$
and \\* $d_{i}(z^{i}_{2s}, z^{i}_{2s+1}) \geq
\alpha(\epsilon)$ for all $s \leq k^{i}-1$;
\item $D_{i}(z^{i}_{2s-1}, z^{i}_{2s}) < 2 \epsilon$.
\end{itemize}
Hence, the sequence $\{\psi_{i}(z^{i}_{s})\}_{s=0}^{2k_{i}-1}$ satisfies:
\begin{itemize}
\item $D(\psi_{i}(z^{i}_{0}),p), D(\psi_{i}(z^{i}_{2k_{i}-1}),q) <
\alpha(\epsilon)$;
\item $\frac{1}{2}\,D(p,q) - 7\epsilon \leq D(\psi_{i}(z^{i}_{2s}),
\psi_{i}(z^{i}_{2s+1})) \leq \frac{1}{2}\,D(p,q) + 8\epsilon$ for $s \leq
k^{i}-2$, \\* $\frac{1}{4}\,D(p,q) -6 \epsilon < D(\psi_{i}(z^{i}_{2k^{i}-2}),
\psi_{i}(z^{i}_{2k^{i}-1})) < \frac{3}{4}\,D(p,q) + 10\epsilon$ and \\*
$d_{i}(\psi_{i}(z^{i}_{2s}), \psi_{i}(z^{i}_{2s+1})) > 0$ for all $s \leq
k^{i}-1$;
\item $D(\psi_{i}(z^{i}_{2s-1}), \psi_{i}(z^{i}_{2s})) < 6 \epsilon$.
\end{itemize}
Hence, for every $n$ such that $\epsilon_{n} < \alpha(d(p,q)/8))$ we can find a sequence
$\{\alpha^{n}_{s}\}_{s=0}^{2k_{n}-1}$ in $\mathcal{M}$ satisfying the above
properties.\footnote{In the sequel, the reader
should keep in mind that these finite sequences can be extended to infinite
ones by setting every element after $2k_{n}-1$ equal to $q$.}  By using a
diagonalisation argument, we can find a subsequence (which we label with the
same index) such that $k_{n+1} \geq k_{n}$ for all $n \in \NN$ and a sequence
$\{\alpha_{s}\}_{s=0}^{2\sup_{n}k_{n} -1}$ such that $\alpha^{n}_{s} 
\stackrel{n
\rightarrow \infty}{\rightarrow} \alpha_{s}$ for all $s \leq 
2\sup_{n}k_{n} -1$.
Obviously $\sup_{n}k_{n}$ must be finite, since otherwise we would 
have found an
infinite sequence of points which are all a distance greater or equal than
$\frac{1}{2}\,D(p,q)$ apart, which is impossible by compactness.\footnote{To
see this, the reader should use the fact that the strong distance is increasing
along causal paths.}  Hence, we have found a finite sequence of points $\beta_s
\leq \beta_{s+1}$, $s=0 \ldots k$, such that:
\begin{itemize}
\item $\beta_{0}=p$ and $\beta_{k} = q\;$;
\item $\sum_{s=0}^{k-1}d(\beta_{s}, \beta_{s+1}) = d(p,q)\;$;
\item $D(\beta_{s},\beta_{s+1}) = \frac{1}{2}\,D(p,q)$, $s \leq k-2$ and
\\*
$\frac{1}{4}\,D(p,q) \leq D(\beta_{k-1},q) \leq \frac{3}{4}\,D(p,q)$.
\end{itemize}
It is possible that for some $s$, $d(\beta_{s}, \beta_{s+1}) = 0$, but
these are limits of timelike intervals as follows from the construction.
Subdividing each of these approximating timelike intervals and using a
compactness argument together with the continuity of $K$ in the strong
topology, one obtains that every two timelike related points are connected
by a causal geodesic. \hfill$\square$
\\*
\\*
This result is, in the authors' viewpoint, very encouraging since it shows that
all concepts fit nicely together.  Notice also that the proof is considerably
more difficult than the one in the metric case where it suffices to use the
existence of a midpoint for path metrics.

\section{Compactness of classes of Lorentz spaces}

In this section, we give some criteria for a collection of Lorentz spaces to
be precompact with respect to the generalized Gromov-Hausdorff uniformity.
The motivation for such criteria is that they give metric-type conditions
under which a class of Lorentz spaces is ``bounded" in some sense, which
is desirable because it makes the class more controllable; an example would
be the possibility of defining summations over classes of discrete 
Lorentz spaces
which are known to be bounded when formulating a finite quantum 
dynamics for causal
sets. The ideas presented here can be traced back to Gromov, and proofs of the
results at hand can be found in Petersen \cite{Pete}.

Let $(\mathcal{M},d)$ be a Lorentz space, and  define (as in Gromov 
\cite{Gromov})

\begin{itemize}

\item $\textrm{Cap}_{\mathcal{M}} ( \epsilon) = $ maximum number
of disjoint ${\epsilon\over2}$-balls in $(\mathcal{M},D)$.

\item $\textrm{Cov}_{\mathcal{M}} (\epsilon) = $ minimum number of
$\epsilon$-balls needed to cover $\mathcal{M}$.

\end{itemize}

Clearly, $\textrm{Cov}_{\mathcal{M}} (\epsilon ) \leq
\textrm{Cap}_{\mathcal{M}}
(\epsilon)$ and both are decreasing functions of $\epsilon$.  What do these
definitions mean?  $\textrm{Cov}_{\mathcal{M}} (\epsilon)$ tells us that one
can choose $\textrm{Cov}_{\mathcal{M}} (\epsilon)$ points $p_{i}$ in
$\mathcal{M}$ such that the pair $\left( \left\{ p_{i} \mid i = 1 \ldots
\textrm{Cov}_{\mathcal{M}} (\epsilon)\right\},d\right)$ is $(2 \epsilon,
\epsilon)$-close in the Gromov-Hausdorff sense to $(\mathcal{M},d)$.  On the
other hand, suppose that $(\mathcal{M}_{1},d_{1})$ and $(\mathcal{M}_{2} ,
d_{2})$ are $(\epsilon, \delta)$ Gromov-Hausdorff close; then we know from
theorem 2 in \cite{Noldus2} that
$$
        d_{\rm GH} ( (\mathcal{M}_{1},D_{1}),
        (\mathcal{M}_{2},D_{2})) \leq \epsilon + 
{\textstyle\frac{3\delta}{2}}\;,
$$
and therefore one obtains from the triangle inequality that
$$
        \textrm{Cov}_{\mathcal{M}_{1}} ( \gamma + 2 \epsilon + 3 \delta)
        \leq  \textrm{Cov}_{\mathcal{M}_{2}} ( \gamma )
$$
and
$$
        \textrm{Cap}_{\mathcal{M}_{1}} (\gamma )
        \geq \textrm{Cap}_{\mathcal{M}_{2}} ( \gamma + 4 \epsilon + 6 \delta )
$$
for all $\gamma > 0$.  Since we have a quantitative Hausdorff uniformity on
$\mathcal{LS}$ with a countable basis around every point, the following two
criteria for compactness are equivalent:

\begin{itemize}

\item Every open cover has a finite subcover.

\item Every sequence has a subsequence which converges to a limit point.

\end{itemize}

\begin{theo} For a class $\mathcal{C} \subset \mathcal{LS}$, the
following statements
are equivalent:

\begin{enumerate}

\item $\mathcal{C}$ is precompact in $\mathcal{LS}$, i.e., every sequence
in $\mathcal C$ has a subsequence that is convergent in $\mathcal{LS}$;

\item There is a function $N( \epsilon ): (0 ,\alpha ) \rightarrow
(0, \infty)$ such that $\textrm{Cap}_{\mathcal{M}} ( \epsilon ) \leq N(
\epsilon )$ for all $(\mathcal{M},d) \in \mathcal{C}$;

\item There is a function $N( \epsilon ): (0 ,\alpha ) \rightarrow
(0, \infty)$ such that $\textrm{Cov}_{\mathcal{M}} ( \epsilon ) \leq N(
\epsilon )$ for all $(\mathcal{M},d) \in \mathcal{C}$.

\end{enumerate}

\end{theo}

\noindent\textsl{Proof:}
\\*
\noindent $1 \Rightarrow 2)$: If $\mathcal{C}$ is precompact, then
for any $\epsilon > 0$ there exist points $(\mathcal{M}_{1},d_{1})$, ...,
$(\mathcal{M}_{k},d_{k}) \in \mathcal{C}$ such that any $(\mathcal{M},d)$ is
$(\frac{\epsilon}{16}, \frac{\epsilon}{24})$ close to some
$(\mathcal{M}_{i},d_{i})$.  Hence, $\textrm{Cap}_{\mathcal{M}}(\epsilon) \leq
\textrm{Cap}_{\mathcal{M}_{i}} (\frac{\epsilon}{2}) \leq \max_{j}
\textrm{Cap}_{\mathcal{M}_{j}} (\frac{\epsilon}{2})$, which clearly proves a
bound for $\textrm{Cap}_{\mathcal{M}}(\epsilon)$ for any $\epsilon > 0$. \\*

\noindent $2 \Rightarrow 3)$ is obvious. \\*

\noindent $3 \Rightarrow 1)$: Because of the generalized triangle inequality,
it suffices to show that for any $\epsilon >0$ there exists a finite collection
$\mathcal{A}$ of spaces in $\mathcal{LS}$ such that any pair $(\mathcal{M},d)
\in \mathcal{C}$ is $(\epsilon, \epsilon)$-close to one of the elements in
$\mathcal{A}$.  Observe that for any $(\mathcal{M},d)$ and $\delta > 0$:
$tdiam(\mathcal{M}) \leq 2 \delta\, \textrm{Cov}_{\mathcal{M}} (\delta)$
since $D_{\mathcal{M}}(p,q) \geq d(p,q)$ for all $p,q \in \mathcal{M}$.
The hypothesis implies the existence of a  function $N(\epsilon)$
such that $\textrm{Cov}_{\mathcal{M}}(\frac{\epsilon}{8}) \leq
N(\frac{\epsilon}{8})$.  Hence every space in $\mathcal{C}$ is
$(\frac{\epsilon}{4}, \frac{\epsilon}{8})$-close to a finite space with
$N(\frac{\epsilon}{8})$ elements, such that the timelike distance between
any two points does not exceed the value $\frac{\epsilon}{4}\,
N(\frac{\epsilon}{8})$.  The Lorentz metric on such a finite space  consists
of a square matrix $(d_{ij})_{1 \leq i,j \leq  N(\epsilon/8)}$ such that $0
\leq  d_{ij} \leq \frac{\epsilon}{4}\, N(\frac{\epsilon}{8})$.  Obviously, one
can find a finite collection $\mathcal{A}$ of Lorentzian metric spaces with
$N(\frac{\epsilon}{8})$ elements such that any of the $(d_{ij})_{1 \leq i,j
\leq  N(\epsilon/8)}$ is $(\frac{\epsilon}{4}, 0)$-close to some element of
$\mathcal{A}$.  Hence, all spaces $(\mathcal{M},d) \in \mathcal{C}$ are
$(\frac{\epsilon}{2}, \frac{5\epsilon}{8})$-close to some element of
$\mathcal{A}$ which concludes the proof. \hfill$\square$ \\*

We show that the covering property with covering function $N$ is stable under
generalized Gromov-Hausdorff convergence provided that $N$ is
continuous (cf.\ the $\mathcal{C}^{\pm}_{\alpha}$ properties in Ref
\cite{Noldus2}).

\begin{theo}

Let $\mathcal{C}(N(\epsilon))$ be the collection of pairs $(\mathcal{M},d) \in
\mathcal{LS}$ such that $\textrm{Cov}_{\mathcal{M}}(\epsilon) \leq
N(\epsilon)$ for all $\epsilon > 0$; suppose $N$ is continuous. Then
$\mathcal{C}(N(\epsilon))$ is compact.

\end{theo}

\noindent\textsl{Proof:}
\\*
\noindent We already know that $\mathcal{C}(N(\epsilon))$ is precompact, hence
suppose $(\mathcal{M}_{i},d_{i}) \stackrel{i \rightarrow \infty}{\rightarrow}
(\mathcal{M},d)$ in the generalized Gromov-Hausdorff uniformity, then with
$\alpha_{i} \stackrel{i \rightarrow \infty}{\rightarrow} 0$ such that
$(\mathcal{M}_{i},d_{i})$ and $(\mathcal{M},d)$ are $(\alpha_{i},
\alpha_{i})$-close, we obtain that
$$
     \textrm{Cov}_{ \mathcal{M}}(\epsilon) \leq \textrm{Cov}_{\mathcal{M}_{i}}
     ( \epsilon - 5 \alpha_{i} )  \leq N( \epsilon - 5 \alpha_{i})\;.
$$
The continuity of $N$ concludes the proof. \hfill$\square$

\section*{Acknowledgements}
JN wishes to thank Norbert Van den Bergh and Frans Cantrijn for continuous
encouragement and Benny Malengier for showing him how to make \TeX\
pictures.  JN also thanks LB for the nice stay in Mississippi during 
the summer.  This work was supported in part by NSF grant
number PHY-0010061 to the University of Misssissippi.

\section*{Appendix A}
In this Appendix, we prove Theorem 2.  First, we introduce some notational
conventions.  Denote by $\alpha^{i}_{j,k}$ the $k$-th element of the $j$-th
column $K^{i}_{j}$ in the causal set $\mathcal{P}^{L}_{i}$.  The labeling of
elements in a column starts from zero.  For example: the maximal element in
$K^{1}_{2}$ is $\alpha^{1}_{2,L+1}$.  For notational simplicity, we agree that
$\alpha^{i}_{j,0} \equiv b^{i}_{j}$ and the top element of the column
$K^{i}_{j}$ is denoted by $t^{i}_{j}$.  In $\mathcal{P}_{2}^{L}$, this results
in $t^{2}_{j} = \alpha^{2}_{j, L+2-j}$.  The idea of the proof is to determine
how the bottom and top elements shift under the maps $\psi$ and $\zeta$.  The
following Lemma is crucial.
\newtheorem{lem}{Lemma}
\begin{lem}
Let $r < \frac{L}{4}+5$ ($1 < r <\frac{L}{4}+5$), and suppose $\zeta (
t^{2}_{r} ) \in  K^{1}_{s} $ ($\psi ( t^{1}_{r} ) \in  K^{2}_{s}$); then $\zeta
( b^{2}_{r+j}) \in \left\{ b^{1}_{s-1}, b^{1}_{s}, b^{1}_{s+1} \right\}$
($\psi ( b^{1}_{r}) \in \left\{ b^{2}_{s-1}, b^{2}_{s}, b^{2}_{s+1} \right\}$),
where $j=0,1$ and all the indices have to be taken modulo $L+1$.
\end{lem}
\textsl{Proof:}
\\*
Remark first that $\zeta ( t^{2}_{r} ) \in K^{1}_{s}$
($\psi ( t^{1}_{r} ) \in  K^{2}_{s}$) with $s$ a natural
number between $2$ ($1$) and $r+k$ if $r < \frac{L}{2} + 1 - k$ ($r \geq
\frac{L}{2} + 1 - k$).  Obviously, $\zeta ( t^{2}_{r} ) \geq \alpha^{1}_{s,
L+2-r-k}$ where $\geq$ means ``in the causal future of".  Suppose $\zeta (
b^{2}_{r+j}) \notin \left\{ b^{1}_{s-1}, b^{1}_{s}, b^{1}_{s+1} \right\}$ for
some $j=0,1$; then $\zeta ( b^{2}_{r+j}) =  \alpha^{1}_{r,q}$ with $q \geq 1$
since
$$
     d_{2}(b^{2}_{r+j}, t^{2}_{r}) - k \geq L + 2
     - (\textstyle{\frac{L}{4}} +4) - (\textstyle{\frac{L}{4}}-1)
     = \textstyle{\frac{L}{2}} - 1 > 0\;.
$$
But in this case $\zeta (t^{2}_{r+1}) \geq \alpha^{1}_{s,L+1-k-r+q}$. 
The above
calculation reveals that $d_{2}(b^{2}_{r+2},t^{2}_{r+1})-k \geq
\frac{L}{2}-2 > 0$ and since moreover $d_{2}(b^{2}_{r+j}, b^{2}_{r+2}) = 0$,
we obtain that $\zeta ( b^{2}_{r+2}) \leq \alpha^{1}_{s, q + j}$.  Hence
$$
     d_{1}(\zeta ( b^{2}_{r+2}),\zeta ( t^{2}_{r}))
     \geq L + 2 - k - r + q - (k + q) \geq \frac{L}{4}\;,
$$
which is impossible since $d_{2}(b^{2}_{r+2}, t^{2}_{r}) = 0$.  The result for
$\psi$ is obvious. \hfill$\square$
\\*
\\*
We shall further construct $\zeta$ and state similar properties of $\psi$
later on.
\begin{lem}
$\zeta (b^{2}_{r}) = b^{1}_{s}$ if $\zeta(t^{2}_{r}) \in K^{1}_{s}$ with $r$
between $1$ and $\frac{L}{4}+3$.
\end{lem}
\textsl{Proof:}
\\*
According to Lemma $1$, we have that $\zeta(b^{2}_{r}) \in
\left\{b^{1}_{s-1}, b^{1}_{s} , b^{1}_{s+1} \right\}$.  Suppose that
$\zeta(b^{2}_{r}) = b^{1}_{s+1}$; then we show that $\zeta(b^{2}_{r+1}) \notin
\left\{b^{1}_{s-1}, b^{1}_{s} , b^{1}_{s+1} \right\}$ which is impossible by
the same Lemma.  The arguments for $\zeta(b^{2}_{r}) = b^{1}_{s-1}$ are
identical.  Suppose $\zeta(b^{2}_{r+1}) = b^{1}_{s+1}$; then $d_{1}(
\zeta(b^{2}_{r}) , \zeta(t^{2}_{r+2})) \geq L+2-(r+2)-k \geq \frac{L}{2} - 2
\geq \frac{L}{4}$ which is impossible since $d_{2} ( b^{2}_{r} , t^{2}_{r+2})
= 0$. Hence suppose that $\zeta (b^{2}_{r+1}) = b^{1}_{s-1}$, then $\zeta
(t^{2}_{r+1}) \in K^{1}_{s}$.  Hence $\zeta(b^{2}_{r+2}) \in \left\{
b^{1}_{s-1}, b^{1}_{s} , b^{1}_{s+1} \right\}$ which is impossible since then
$d_{1}(\zeta(b^{2}_{r+2}) , \zeta(t^{2}_{r}) \geq L+2-r-k \geq \frac{L}{2}$.
So, we are only left with $\zeta(b^{2}_{r+1}) = b^{1}_{s}$.  Obviously $\zeta
(t^{2}_{r+1}) \in K^{1}_{s+1}$, since otherwhise $\zeta (t^{2}_{r+1}) \in
K^{1}_{s}$ which was proven impossible before.  Hence, $\zeta (b^{2}_{r+2})
\in \left\{ b^{1}_{s}, b^{1}_{s+1}, b^{1}_{s+2} \right\}$.  Previous arguments
show that $\zeta (b^{2}_{r+2}) = b^{1}_{s+2}$, but then $\zeta (t^{2}_{r+2})
\geq \alpha^{1}_{s+1, L-r-k}$ which implies that $d_{1}( \zeta(b^{2}_{r}),
\zeta(t^{2}_{r+2})) \geq \frac{L}{2}-2 \geq \frac{L}{4}$. \hfill$\square$
\\*
\\*
Obviously, the same theorem applies to $\psi$ for $1<r<\frac{L}{4}+4$.
The following Lemma almost gives the necessary result.
\begin{lem}
If $\zeta ( t^{2}_{1}) \in K^{1}_{s}$ with $s$ ranging between $2$ and $k+1
\leq \frac{L}{4}$ then $\zeta (b^{2}_{i}) = b^{1}_{s+i-1}$ and
$\zeta(t^{2}_{i}) \geq \alpha^{1}_{s+i-1, L+2-i-k}$ for $i \leq \frac{L}{4}+3$.
\end{lem}
\textsl{Proof:}
\\*
As a consequence of Lemma 2 we have only two possibilities.  Either $\zeta
(b^{2}_{i}) = b^{1}_{s+i-1}$ and $\zeta(t^{2}_{i}) \geq \alpha^{1}_{s+i-1,
L+2-i-k}$ or $\zeta (b^{2}_{i}) = b^{1}_{s-i-1}$ and $\zeta(t^{2}_{i}) \geq
\alpha^{1}_{s-i-1, L+2-i-k}$ for $i \leq \frac{L}{4}+3$.  If the latter were
true then $\zeta(t^{2}_{s+1}) \geq \alpha^{1}_{l,L+2-s-k}$ since $s+1 \leq
\frac{L}{4} +1$ which is impossible since $ L+2-s-k > 2$. \hfill$\square$
\\*
\\*
First of all, it is easy to see that if $\psi ( t^{1}_{2} ) \in
K^{2}_{\tilde{s}}$ with $\tilde{s}$ between $2$ and $k+1$.  Since suppose
$\psi ( t^{1}_{2} ) \in K^{2}_{1}$ then $\psi(b^{1}_{1}) = b^{2}_{L}$ since
otherwise or $\psi(b^{1}_{1}) \geq \alpha^{2}_{1,1}$ or $\psi(b^{1}_{1}) \in
\left\{ b^{2}_{1}, b^{2}_{2} \right\}$.  The former is impossible since then
$d_{2}( \psi (b^{1}_{3}), \psi (t^{1}_{1}) ) \geq \frac{L}{2} + 2 - k >
\frac{L}{4}$.  The latter would imply that $\psi(b^{1}_{1}), \psi(t^{1}_{3}))
\geq L-1-k$ which is also impossible.  But then $\psi(t^{1}_{1}) \in
K^{2}_{L-1} \cup K^{2}_{L} \cup K^{2}_{1}$, which is impossible for $L \geq 8$
(which was the assumption).  By a reasoning analogous to the one in Lemma 3,
we obtain that $\psi (b^{1}_{i}) = b^{2}_{\tilde{s}+i-2}$ and $\psi(t^{1}_{i})
\geq \alpha^{2}_{\tilde{s}+i-2, L+2-i-k}$ for $1< i \leq \frac{L}{4}+3$.
Moreover, $\psi(b^{1}_{1}) = b^{2}_{\tilde{s}-1}$.
\\*
\\*
We finish the proof by remarking that $d_{1}(b^{1}_{s+j}, \zeta( t^{2}_{1}))
\geq L+1-k$ for $j=-1,0,1$.  Since $1 \geq s-1 <s+1 \leq \frac{L}{4} + 1$, we
have that $\psi (b^{1}_{s+j}) = b^{2}_{s + \tilde{s} - 2 + j}$.  Since $L+1-2k
\geq \frac{L}{2} + 3$ we obtain that $\psi \circ \zeta (t^{2}_{1}) \in
K^{2}_{s + \tilde{s} - 2}$.  Hence $D_{2}(t^{2}_{1}, \psi \circ \zeta
(t^{2}_{1})) = L$ which finishes the proof.  \hfill$\square$

\end{document}